\documentclass[aps,preprint,amsmath,amssymb,groupedaddress,superscriptaddress]{revtex4-1}
\usepackage{graphicx}
\usepackage{float}
\usepackage{longtable}
\usepackage[usenames]{color}
\usepackage[normalem]{ulem}
\usepackage{amsmath}
\usepackage{amssymb}
\usepackage{amsfonts}
\usepackage{dcolumn} 
\usepackage{bm}      
\usepackage{subfigure}
\usepackage{rotating}
\usepackage{fontenc}
\usepackage{cancel}
\usepackage{amsthm}
\usepackage{hyperref}

\begin{document}

\title{Articulation Points in Complex Networks}

\author{Liang Tian}
\affiliation{Channing Division of Network Medicine, Brigham and Women's Hospital and Harvard Medical School, Boston, MA 02115, USA}
\affiliation{College of Science, Nanjing University of Aeronautics and Astronautics, Nanjing 210016, China}
\author{Amir Bashan}
\affiliation{Department of Physics, Bar-Ilan University, Ramat-Gan 52900, Israel} 
\author{Da-Ning Shi}
\affiliation{College of Science, Nanjing University of Aeronautics and Astronautics, Nanjing 210016, China}
\author{Yang-Yu Liu}
\email[Correspondence: ]{yyl@channing.harvard.edu}
\affiliation{Channing Division of Network Medicine, Brigham and Women's Hospital and Harvard Medical School, Boston, MA 02115, USA}
\affiliation{Center for Cancer Systems Biology, Dana Farber Cancer Institute, Boston, MA 02115, USA }

 \date{\today}

\begin{abstract}
An articulation point in a network is a node whose removal disconnects
the network. Those nodes %
play key roles in ensuring connectivity of %
many real-world networks, 
from infrastructure networks %
to %
protein interaction networks and
terrorist communication networks.
Despite their fundamental importance, a general framework of studying
articulation points in complex networks is lacking. Here we develop
analytical tools to study key %
issues pertinent to articulation points, 
e.g. the expected number of them and the
network vulnerability against their removal, in an arbitrary complex
network. 
We find that a greedy articulation point removal process
  provides us a novel perspective on the organizational principles 
of complex networks. Moreover, this process is %
  associated with 
  two
fundamentally different types of 
percolation transitions with a rich phase diagram. 
Our results shed light on the design of more resilient
infrastructure networks and %
the effective destruction of
terrorist communication networks. %
\end{abstract}

\maketitle
\tableofcontents

\section{Introduction}

A fundamental challenge in studying complex networked systems is to
reveal the interplay between network topology and
functionality\cite{albert2002statistical,newman2003structure}. 
Here we tackle this challenge by investigating a
classical notion in graph theory, i.e. articulation points. 
A node in a network is an articulation point (AP) if its removal
disconnects the network or increases the number of connected
components in the network%
\cite{behzad1972introduction, harary6graph}. 
Those AP nodes play important roles in ensuring the connectivity 
of many real-world networks.  
For example, in infrastructure networks such as air traffic networks
or power grids, APs, if disrupted or attacked, pose serious risks to
the infrastructure 
\cite{conti2013disruptions, cox2009architecturally}.
In wireless sensor networks, failures of APs will block data
transmission from one network component to others 
\cite{Hassan}. 
In the yeast protein-protein interaction network, lethal mutations are
enriched in the group of highly connected proteins that are APs 
\cite{prvzulj2004functional}. 
Identification of APs also helps us better solve other challenging problems, e.g. the calculation of determinants of large matrices 
\cite{maybee1989matrices}; and the vertex cover problem on large graphs, which is a notorious NP-complete problem 
\cite{patel2014improved}.

Representing natural
vulnerabilities of a network, 
APs are %
potential
targets of attack if one aims for immediate damage to a network. 
For example, in the %
terrorist communication network of
the 9/11 attacks on U.S. (Fig.1A), cutting off those AP members (shown
in red) could have caused a great deal of disruption in the
terrorists' ability to communicate 
\cite{krebs2002mapping}. 
Note that the removal of an AP in a network may lead to the
emergence of new APs in the remainder of the network.  
A brute-force strategy of network destruction is to \emph{iteratively} remove
the most destructive AP that will cause the most nodes disconnected
from the giant connected component (GCC) of the current network. Given
a limited ``budget'' (i.e. the number of nodes to be removed), this
AP-targeted attack (APTA) strategy is very efficient in reducing the 
GCC, compared with other existing strategies
\cite{albert2000error, cohen2000resilience,
  morone2015influence}. Indeed, we find that for a small fraction of removed nodes APTA leads to the quickest reduction of GCC for a wide range of
real-world networks from technological to infrastructure, biological,
communication, and %
social networks (see Appendix \ref{Sec.SI-I-A}). 
Depending on the initial network structure, APTA
would either completely decompose the network or result in a residual
GCC that occupies a finite fraction of the network. This residual GCC
is a biconnected component (or bicomponent), in which any two nodes
are connected by at least two independent paths and hence no AP exists
\cite{tarjan1985efficient}. 
For accuracy, we will call it residual giant bicomponent (RGB)
hereafter. 
This RGB naturally represents a \emph{core} that maintains the 
\emph{structural integrity} of the network.  

Identification and removal of APs also provide us a novel perspective on
the organizational principles of complex networks.   
For example, in a terrorist communication network,  
each AP member can be considered as a
\emph{messenger} %
of a particular subnetwork, %
because any information exchange between that subnetwork %
and the rest of the network passes through the AP. All the APs and
their associated subnetworks %
in the original network constitute the \emph{first layer} of the
terrorist network. After removing all the APs in the original network,
the first layer is peeled off, new APs emerge and the \emph{second
  layer} of the network is exposed. We can repeat this %
greedy APs removal (GAPR) process until there is no AP left in the
network. Figure 1 illustrates this novel network decomposition
process in the 9/11 terrorist communication network, which has 62 terrorists.  %
We find that this network consists of 3 layers %
and an RGB
of 26 nodes. %
(Note that the RGB associated with the GAPR process is similar to but
not necessarily the same as that of the APTA process, see Appendix \ref{Sec.SI-I-B}) 
Interestingly, among those 26 RGB nodes, 16 of them are hijackers in 
the 9/11 terrorist attack, 
which in total has 19 hijackers. 
In a sense, this RGB serves
as a \emph{core} %
maintaining the {\it functionality} of this covert network, which has %
a particular goal---hijacking. 
Note that
 some of the
hijackers in the RGB are not hubs (i.e. highly connected
nodes), but only have two or three neighbors in the network. Hence they cannot be easily
identified through traditional network decompositions, e.g. the
classical $k$-core decomposition \cite{dorogovtsev2006k,Kitsak-NP-2010}.

Despite the importance of APs in ensuring the connectivity of
many real-world networks, 
we still lack a deep understanding on
the roles and properties of APs in many complex networks. We don't
have a theory to quantify if a given network has overrepresented or
underrepresented APs comparing to its randomized counterparts. Moreover,
the implication of the RGB serving as a core that maintains
the structural integrity %
and/or functionality of a network has never been explored %
before. 
In this article we offer an analytical framework to study those %
fundamental issues pertinent to APs 
in complex networks. %

\section{Results}%
\subsection{Analysis of Real Networks}
We first study the fraction of APs ($n_{\text{AP}}$) and the relative
size of the RGB ($n_{\text{RGB}}$) in a wide range of real-world
networks. 
The APs of a network can be identified all together by a linear-time
algorithm based on depth-first search \cite{tarjan1985efficient}. 
As for RGB, we find that the two RGBs associated with %
the APTA and GAPR processes significantly overlap for many %
real-world networks (Appendix \ref{Sec.SI-I-B}).  
Hereafter %
we focus on the RGB obtained from the GAPR process, which is
deterministic and hence analytically solvable. 

We find that %
many %
real networks have a non-ignorable fraction of APs 
and %
a rather small RGB  (Fig.2A). %
One may expect that infrastructure networks should have a relatively
small fraction of APs and a large RGB, and hence are very robust
against AP removal. 
Ironically, this is not the case. The power
grids in two regions of U.S. have almost the largest fraction ($\sim 24\%$) of APs among all the real networks analyzed in this
work. And %
they have almost no RGB.
The road networks of three states in U.S. have almost 20\% of APs, and
a small RGB %
($ n_\text{RGB} \sim 0.08$).     
In contrast, among all the 28 food webs we analyzed, 22 of them have
no APs  (and hence $n_\text{RGB}=1$), indicating that those networks
are biconnected and the extinction of one species will not disconnect
the whole network.  
More interestingly, 
we find that most of the real networks analyzed here have either a very small RGB or a rather big one 
(Fig. 2A  light red and green regions).
Later we will show that this phenomenon is related to a novel
discontinuous phase transition associated with the GAPR process. 

To identify the topological characteristics that determine these two
quantities ($n_\text{AP}$ and $n_\text{RGB}$), 
we compare $n_\text{AP}$ (or $n_\text{RGB}$) of a given real network with that of its randomized counterpart.
To this aim,
we %
randomize each real network using a complete
randomization procedure that turns the network into an Erd\H os-R\'enyi (ER) type of random network with the number of nodes $N$ and
links $L$ unchanged \cite{erd6s1960evolution}. 
We find that most of the completely randomized networks
possess very different $n_\text{AP}$ (or $n_\text{RGB}$), comparing to their corresponding real networks (Fig. 2B, C). This indicates that %
complete randomization eliminates the topological characteristics that
determine $n_{\text{AP}}$ and $n_{\text{RGB}}$. 
By contrast, when we apply %
a degree-preserving randomization, which
rewires the links %
among nodes while keeping the degree $k$ of
each node unchanged,  %
this procedure does not alter
$n_{\text{AP}}$ and $n_{\text{RGB}}$ significantly (Fig. 2D, E). %
In other words, the characteristics
of a network in terms of $n_{\text{AP}}$ and $n_{\text{RGB}}$ is largely encoded
in its degree distribution $P(k)$. %
Most of the real-world networks display slightly smaller $n_\text{AP}$ and bigger $n_\text{RGB}$ than their degree-preserving randomized counterparts. 
We attribute these differences
to higher-order structure correlations, such
as clustering 
\cite{watts1998collective}  and degree assortativity 
\cite{newman2002assortative}, which are eliminated in the
degree-preserving randomization.
\subsection{Analytical Results}
The results of real-world networks 
prompt us to analytically
calculate %
$n_{\text{AP}}$ %
and $n_{\text{RGB}}$ for networks with any prescribed degree distribution.
This is achieved by analyzing the discrete-time dynamics of the deterministic GAPR process, which generates a series of snapshots for the remainder network with a clear temporal order $\{0,1,...,t,...., T\}$. Here, $T$ is the total number of GAPR steps, which is also the number of layers peeled off during the GAPR process. 
We denote the fraction of APs and the relative size of the GCC in the
original network as $n_{\text{AP}}(0)$ and $n_{\text{GCC}}(0)$,
respectively. Removal of the original APs leads to a new
  fraction of APs $n_{\text{AP}}(1)$ and a smaller GCC of
relative size $n_{\text{GCC}}(1)$. %
We repeat this process and denote the fraction of APs and the relative size of the GCC of the
network snapshot at time step $t$ as $n_{\text{AP}}(t)$ and
$n_{\text{GCC}}(t)$, respectively.  
At the end of the GAPR process, we have $n_\text{AP}(T)=0$ and $n_\text{GCC}(T) = n_\text{RGB}$.
Based on
the configuration model of uncorrelated random networks 
\cite{molloy1995critical, callaway2000network, newman2001random},
we can analytically calculate  $n_{\text{AP}}(t)$ and
$n_{\text{GCC}}(t)$ for networks with arbitrary degree distributions
at any time step $t$. This enables us to further compute $T$ and $n_\text{RGB}$ (see Appendix \ref{Sec.SI-II} for
details). %

\subsubsection{Fraction of APs}
We first investigate the fraction of APs in the original network, i.e. $n_\text{AP}=n_\text{AP}(0)$. 
We calculate $n_\text{AP}$ in two canonical model networks: (1) ER random networks with Poisson degree distributions  $P(k)=e^{-c} c^k/k!$, where $c$ is the mean degree; and (2) scale-free (SF) networks with power-law degree distributions $P(k) \sim k^{-\lambda}$, where $\lambda$ is often called the degree exponent (Fig. 3A-B). 
The fraction of APs 
is trivially zero in the two limits $c\to 0$ and $c\to \infty$, and reaches
its maximum %
at a particular mean degree $c_\text{AP}$. 
For ER networks, 
we find that $c_\text{AP}=1.41868\cdots$,
which is larger than $c_\text{p}=1$, the
critical point of ordinary percolation
where the GCC emerges.

The phenomenon that $n_\text{AP}$ displays a unimodal behavior and the fact that $c_\text{AP} > c_\text{p}$ can be explained as follows. 
The process of increasing the mean degree $c$ of an ER network can be considered as the process of randomly adding links into the network.
When the mean degree $c$ is very small (nearly zero), there are only
isolated nodes and dimers (i.e. components consisting of two nodes
connected by one link), and thus $n_{\text{AP}}\rightarrow 0$. 
With $c$ gradually increasing but still smaller than $c_{\text{p}}$,
the network is full of %
finite connected components (FCCs), most of which are trees (Fig. 3C). 
Hence, in the range of $0<c<c_{\text{p}}$ most of the
nodes (except isolated nodes and leaf nodes) are APs, and adding more
links to the network will decrease the number of isolated nodes and
dimers and hence increase the number of APs. 
 When $c>c_{\text{p}}$, 
the GCC develops and occupies a finite fraction of nodes in the
network (highlighted in light blue in Fig. 3D-F).  
In this case, %
we can classify the links to be added to the network into two types: %
(I) links inside the %
GCC (yellow dashed lines);
and 
(II) links that 
 connect the 
GCC with an FCC 
or connect two %
FCCs (green dashed lines). 
The probability that an added link is type-I (or type-II) as a function of the mean degree $c$ is shown in Fig. 3A (light blue region). 
Adding type-I links to the network 
will never induce 
new APs, and may even convert the %
existing 
APs (see Fig.3D, E, nodes in black boxes) back to normal
nodes. 
By contrast, %
adding type-II links 
will never decrease the number of APs
and could convert normal nodes (see Fig.3C-E, nodes in orange boxes) to APs. 
The contributions of these two %
types of links to $n_{\text{AP}}$ compete
with each other. 
At the initial stage of this range of $c$ ($c > c_\text{p}$), 
since the GCC is still small, most of
the added links are type-II (Fig. 3A, green dashed line), %
and thus $n_{\text{AP}}$ continues to
increase (Fig. 3A, red line). 
At certain point $c_\text{AP}$, where the peak of
$n_{\text{AP}}$ locates, %
the contribution of type-I links to $n_\text{AP}$ overwhelms that of type-II links, 
hence $n_{\text{AP}}$ begins to decrease. 
When the mean degree $c$ is
large enough, the network itself becomes a bicomponent without any AP. 

The phenomena that $c_\text{AP}  > c_\text{p}$ is even more prominent for SF networks, where $c_\text{p}(\lambda) < 1$ and $c_\text{AP}(\lambda) > 1.41868\cdots$. This is because,
for SF networks, even though the GCC emerges at lower $c_{\text{p}}$, its relative size is rather small at the initial stage of its emergence, which results in larger $c_{\text{AP}}$ (Fig. 3B).

\subsubsection{Two Percolation Transitions of Totally Different Nature}
We now systematically study the behaviors of $n_{\text{GCC}}(t)$ and $n_{\text{RGB}}$ as functions of the mean degree $c$ for
infinitely large ER networks. 
To emphasize the $c$-dependence, hereafter we denote
$n_{\text{GCC}}(t)$ and $n_{\text{RGB}}$ of ER networks as
$n_{\text{GCC}}(t,c)$ and $n_{\text{RGB}}(c)$, respectively.

As shown in Fig. 4A (thin grey lines), after any {\it finite} steps 
of GAPR, the GCC always emerges in a continuous manner,
suggesting a {\it continuous} phase transition. 
Hereafter we will call it GCC percolation transition. 
For $t$ steps of GAPR, the GCC percolation transition %
displays a critical phenomenon: $n_{\text{GCC}}(t,c) \sim
(c-c^*(t))^{\beta (t)}$ for $(c- c^*(t)) \to 0^+$, where $c^*(t)$ is
the critical mean degree, i.e. the percolation threshold, and
$\beta(t)$ is the critical exponent. %
We find that as $t$ increases, $c^*(t)$ becomes larger and larger,
but eventually converges to $c^*(\infty)=c^*=3.39807\cdots$.
Note that, for any finite $t$, the critical exponent $\beta(t)$ associated with the GCC percolation transition is the same: $\beta(t)=\beta_\text{GCC}=1$    
(see Fig. 4C, thin grey lines).

By contrast, if we allow for {\it infinite} steps of GAPR (i.e. we stop the
process only if there is no AP left), the size of the resulting GCC
(which is nothing but the RGB), denoted as $n_\text{RGB}(c)$, displays a
remarkable \emph{discontinuous} phase transition: %
$n_\text{RGB}(c)$ abruptly jumps from zero (when $c<c^*$) to a finite
value at $c^{*}$, and then increases with increasing $c$ (see Fig.4A
thick black line). Hereafter we will call it RGB percolation transition. 
If we denote the jump size as $\Delta$, we find that $n_\text{RGB}(c) - \Delta \sim
(c-c^*)^{\beta_\text{RGB}}$ with critical exponent $\beta_\text{RGB} = 1/2$ when $(c- c^*) \to 0^+$ (Fig. 4C, thick black line), 
suggesting that the RGB percolation transition is actually a {\it hybrid}
phase transition \cite{dorogovtsev2006k, parisi2008k, dorogovtsev2008critical}. 
In other words, $n_\text{RGB}(c)$ has a jump at the critical point $c^*$
as a first-order phase transition but also has a critical singularity
as a second-order phase transition. 
Interestingly, the GCC and RGB percolation transitions have
  completely different critical exponents associated with their
  critical singularities (Fig. 4C).

We also calculate the total number of GAPR steps $T(c)$ needed to
remove all APs of an infinitely large ER network of mean degree $c$ (Fig. 4B).
We find that %
$T(c)$ is %
finite for $c<c^*$; %
diverges when $(c-c^*)\to 0^-$; and is infinite for any $c>c^*$. The divergence of $T(c)$
displays a scaling behavior $T(c) \sim |c-c^{*}|^{-\gamma_{-}}$ with
critical exponent $\gamma_{-}=1/2$ when $(c-c^*)\to 0^-$ (Fig. 4D, red line).

The nature of the discontinuous RGB percolation transition and
the behavior of $T(c)$ can be revealed by
analyzing the dynamics of the GAPR process. In particular,
we can calculate $n_\text{GCC}(t,c)$, $n_\text{AP}(t,c)$, as well as a key
quantity in the GAPR process, i.e. the average number of newly induced APs per single
AP removal: %
$\eta(t,c) = n_\text{AP}(t,c)/n_\text{AP}(t-1,c)$ for
$t > 0$ and at different mean degrees $c$ (Fig. 4E-G). 

For $c > c^*$ (supercritical region), the fraction of APs exponentially
decays as $n_\text{AP}(t,c) \sim \exp( -t/\widetilde{T}(c) )$ after an initial transient time (Fig. 4F, green lines), where $\widetilde{T}(c)$ is the characteristic time scale. In this region, with increasing $t$, $\eta(t,c)$ quickly
reaches an equilibrium value $\eta(\infty,c)=\exp( -1/\widetilde{T}(c))$, which is
smaller than 1 (Fig. 4G,  green lines). Consequently, $n_\text{GCC}(t,c)$
converges to a finite value for $t \to \infty$ (Fig. 4E green lines),
resulting in a finite $n_{\text{RGB}}$. Since $T(c)$ is infinite in this region,
we can use $\widetilde{T}(c)$ to characterize the relaxation behavior of GAPR
process. We find that $\widetilde{T}(c)$ increases as $c$ decreases
(Fig. 4B, green line), and diverges as $\widetilde{T}(c) \sim
|c-c^{*}|^{-\gamma_{+}}$ with critical exponent $\gamma_{+}=1/2$ when $(c-c^*) \to 0^+$ (Fig.4D, green line).  
Note that as $c$ decreases and approaches
$c^*$ from above, the equilibrium value $\eta(\infty,c)$ gradually approaches 1 (Fig.4G).

When $(c - c^*) \to 0^+$ (i.e. right above the criticality), the fraction of APs decays in a
power-law manner for large $t$, i.e. $n_\text{AP} \sim t^{-z}$ with $z=2$ (Fig. 4F and inset, black line), rendering $\eta(\infty,c)=1$ (Fig.4G, black thick line). 
Consequently, $n_\text{GCC}(t,c)$ converges to a finite value in the $t \to
\infty$ limit (Fig. 4E, black thick lines), leading to a finite $n_{\text{RGB}}$. The
fact $\eta(\infty,c)=1$ suggests that in average every removed AP will induce %
one new AP at the next time step, and hence the GAPR
process will continue forever. This explains why $T(c^*)$ diverges. Note that the equilibrium value $\eta(\infty,c)=1$ can only
be reached when $n_\text{AP}(t,c)$ displays a power-law decay as $t \to
\infty$. This is because, as long as $n_\text{AP}(t,c)$ is finite at any
finite $t$, the GAPR will gradually dilute the network, rendering a
larger and larger value of $\eta(t,c)$ as $t$ grows.
When $(c - c^*) \to 0^-$ (i.e. right below the criticality), as well
as in the entire subcritical region ($c<c^*$), 
after an initial decay, $n_\text{AP}(t,c)$ begins to exponentially grow
with increasing $t$ (Fig. 4F, red lines). Consequently, $\eta(t,c)$ is initially less than
1, but then becomes drastically larger than 1 (Fig. 4G, red lines),
which causes $n_\text{GCC}(t,c)$ quickly decays to zero (Fig. 4E, red
lines), and hence $T(c)$ is finite and the RGB dose not exist. 
The sudden collapse of the RGB upon an infinitesimal decrease in $c$ suggests
the discontinuous nature of the RGB percolation transition in ER
networks.    
Note that at time step $T$, the network will break into pieces and
there is no AP left. %
Hence in the last few GAPR steps the growth of $n_\text{AP}(t,c)$ will
slow down and eventually decrease (Fig. 4F, G, tails in the red
lines). 

For finite-size networks sampled from a network ensemble
  with a prescribed degree distribution, the value of $n_{\text{RGB}}$ at criticality $c^{*}$
  is subject to large sample-to-sample fluctuations, %
  being either zero or a large finite value (Fig. 5A and inset), which
  is another evidence of discontinuous phase transition
  \cite{buldyrev2010catastrophic}. 
This discontinuous phase transition also partially explains the fact that real-world networks have either a very small or a rather big RGB (Fig. 2A).

\subsubsection{Phase Diagram}

The nature of the RGB percolation transition %
in SF networks is qualitatively the same as that in ER networks. The transition from the non-RGB phase to the RGB
phase is discontinuous (Fig. 5A).
The critical point $c^*(\lambda)$ increases with
decreasing $\lambda$. Also, the jump of the RGB relative size at criticality
increases as $\lambda$ approaches 2 (Fig. 5A).

The $c-\lambda$ phase
diagram of SF networks is shown in Fig. 5B. The whole diagram
consists of three phases. For $c<c_\text{p}(\lambda)$ (grey region)
, there exists no GCC in the network, and hence no
RGB. For $c_p(\lambda)<c<c^{*}(\lambda)$ (light blue region), even though the GCC may
survive after certain finite steps of GAPR, the RGB still does not
exist. Since in both regimes there is no RGB, we call them non-RGB-I
phase and non-RGB-II phase, respectively. The transition between these
two phases is the ordinary GCC percolation transition, which is
continuous (thick dashed line). Note that, in non-RGB-II phase, the phase transition
associated with the emergence the GCC after any finite $t$ steps of GAPR is still continuous (thin dot-dashed lines). 
For $c>c^*(\lambda)$ (light yellow region), the
network suddenly has an RGB. This regime is
referred to as the RGB phase.
As we mentioned above, the transition between the non-RGB-II
phase and the RGB phase is discontinuous (thick solid line). 
We have performed extensive numerical simulations to confirm our
analytical results (see Appendix \ref{Sec.SI-III} for details).

\section{Discussion}
Network science has been tremendously developed in the last decade and significantly advanced our understanding of many complex systems 
\cite{albert2002statistical, newman2003structure, dorogovtsev2008critical}. 
In particular, structural transitions in complex networks have been extensively studied and found to affect many network properties 
\cite{albert2000error, cohen2000resilience, pastor2001epidemic, palla2005uncovering, derenyi2005clique, dorogovtsev2006k, achlioptas2009explosive, ziff2009explosive,
da2010explosive, grassberger2011explosive, riordan2011explosive, cho2013avoiding, d2015anomalous, buldyrev2010catastrophic, karrer2014percolation, morone2015influence, liu2011controllability, liu2012core}. 
However, some basic graph theoretical concepts are still not fully understood in the context of complex networks, hindering the further development of network science. Articulation point is one of them.
In this article, %
we propose an analytical framework to systematically investigate AP-related issues in complex networks. Many interesting phenomena of APs are discovered and explained for the first time. 
In particular, the novel network decomposition method based on the GAPR process and the concept of the RGB offer us a brand-new perspective on the organizational principles of complex networks.
The presented results
shed light on the design of more resilient infrastructure networks and
more effective destructions of malicious networks such as the criminal
or terrorist communications networks 
\cite{firmani2014not}.

The framework presented here also raises a number of questions, answers to
which could further deepen our understanding %
of complex networks. For example, although our analytical work focuses on
uncorrelated networks, all the quantities considered here can be
efficiently calculated for large real-world  networks, providing a platform to
systematically address the role of higher-order structure correlations on the fraction of AP
nodes and the relative size of the RGB \cite{Maslov-Science-2002, newman2002assortative,
  Pastor-Satorras-PRL-2001}.  
Moreover, the functionalities of the RGB nodes in different real-world
networks can be %
further studied and related to the nature of those networks. %
Taken together, our results indicate that many aspects of articulation
points can be systematically investigated, opening new
avenues to deepening our understanding of complex networked systems.

\newpage
\begin{figure}
\includegraphics[width=\textwidth]{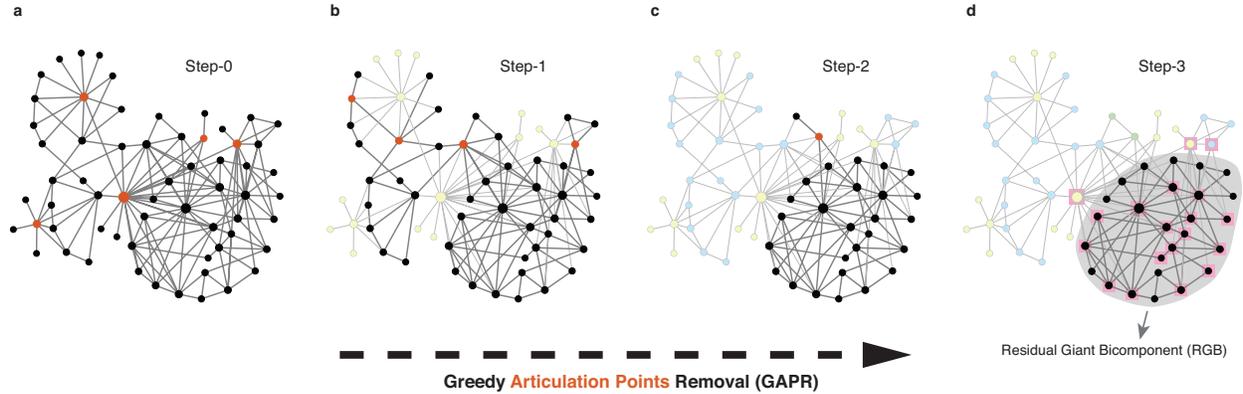}
\caption{{\bf Articulation points and the greedy articulation points removal
process.} ({\bf a}) Articulation points (APs, shown in red) in the
terrorist communications network from the attacks on the United States
on September 11, 2001. This network contains in total 62
  nodes and 153 links \cite{krebs2002mapping}. ({\bf b}-{\bf c}) At each time step, all the APs and the
links attached to them are removed from the network. This greedy
APs removal (GAPR) procedure
can be considered as a novel network decomposition process: at each 
step, all the removed nodes (due to the removal of APs in the current
network) form a layer in the network. And we peel the network off
one layer after another, until there is no AP left. 
We find that this terrorist communications network consists of 3 layers, shown in light yellow,
blue, and green, respectively.
({\bf
  d}) After 3 steps, a well-defined residual giant bicomponent (RGB)
is left, which contains 26 of the 62 nodes. Interestingly, 16 of the 19
  hijackers (highlighted in pink squares) are in the RGB, which is
  statistically significant (Fisher's
  exact test yields a two-tailed test p-value $1.13
  \times 10^{-5}$).}
\end{figure}
\clearpage
\noindent 

\begin{figure}
\includegraphics[width=\textwidth]{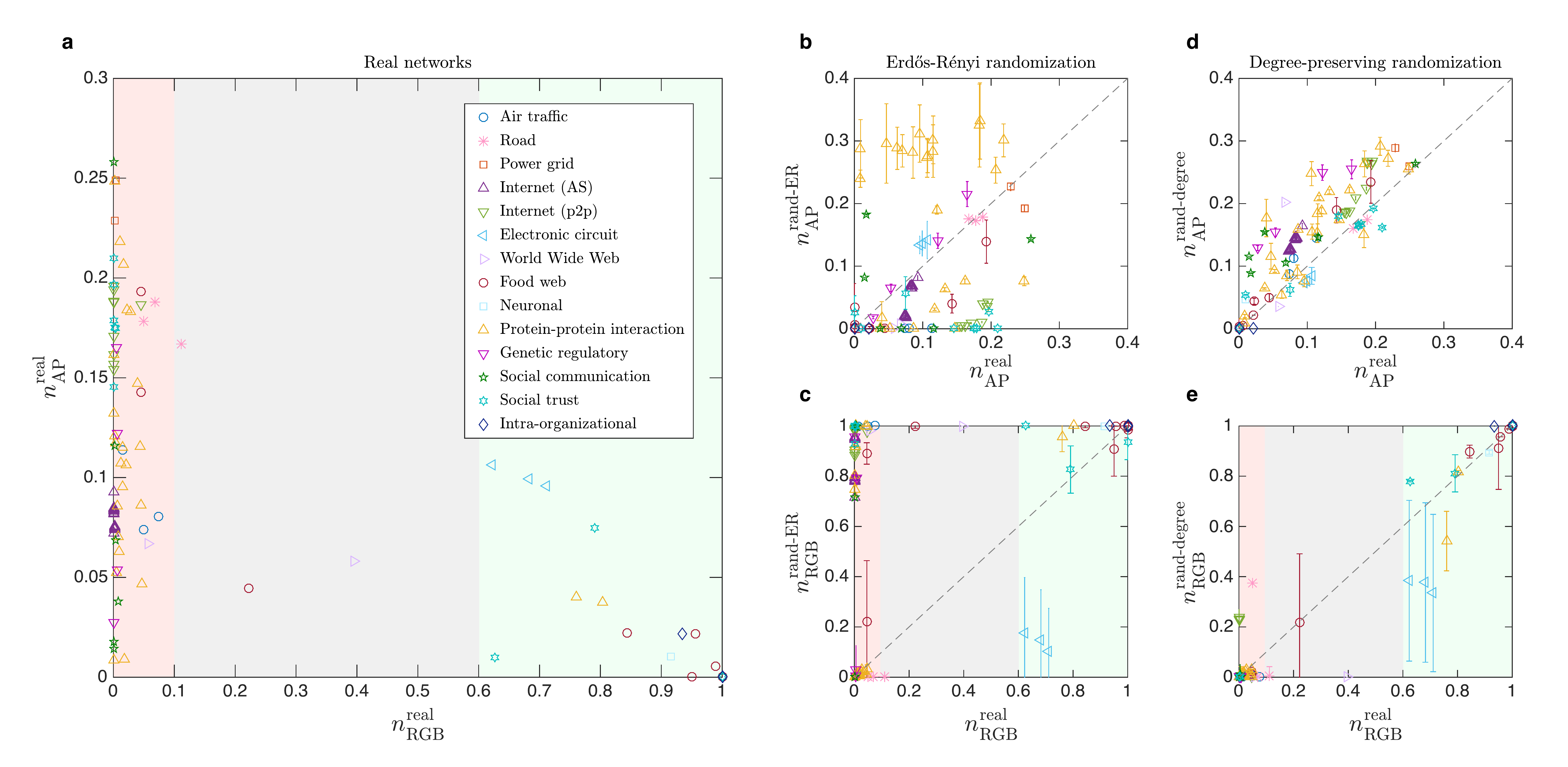}
\caption{{\bf Fraction of articulation points and relative size of the residual
giant bicomponent in real networks and their randomized counterparts.}   
(\textbf{a}) Fraction of APs ($n^{\text{real}}_{\text{AP}}$)
versus relative size of the RGB ($n^{\text{real}}_{\text{RGB}}$) in a wide
range of real networks.  
(\textbf{b}-\textbf{c}) $n^{\text{rand-ER}}_{\text{AP}}$ and
$n^{\text{rand-ER}}_{\text{RGB}}$ calculated from the fully randomized
counterparts of the real networks, compared with the values
($n^{\text{real}}_{\text{AP}}$ and $n^{\text{real}}_{\text{RGB}}$) calculated from the real networks. 
(\textbf{d}-\textbf{e})
$n^{\text{rand-degree}}_{\text{AP}}$ and
$n^{\text{rand-degree}}_{\text{RGB}}$ calculated from the
degree-preserving randomized counterparts of the real networks,
compared with the values calculated from the real networks. 
All data points and error bars (s.e.m.) in \textbf{b}-\textbf{e} are
determined from 100 realizations of the randomized networks. The
dashed lines ($y=x$) are guide for eyes. For detailed description of
these real networks and references, see Appendix \ref{Sec.SI-IV}.
}
\end{figure}
\clearpage
\clearpage
\begin{figure}
\centering\includegraphics[width=\textwidth]{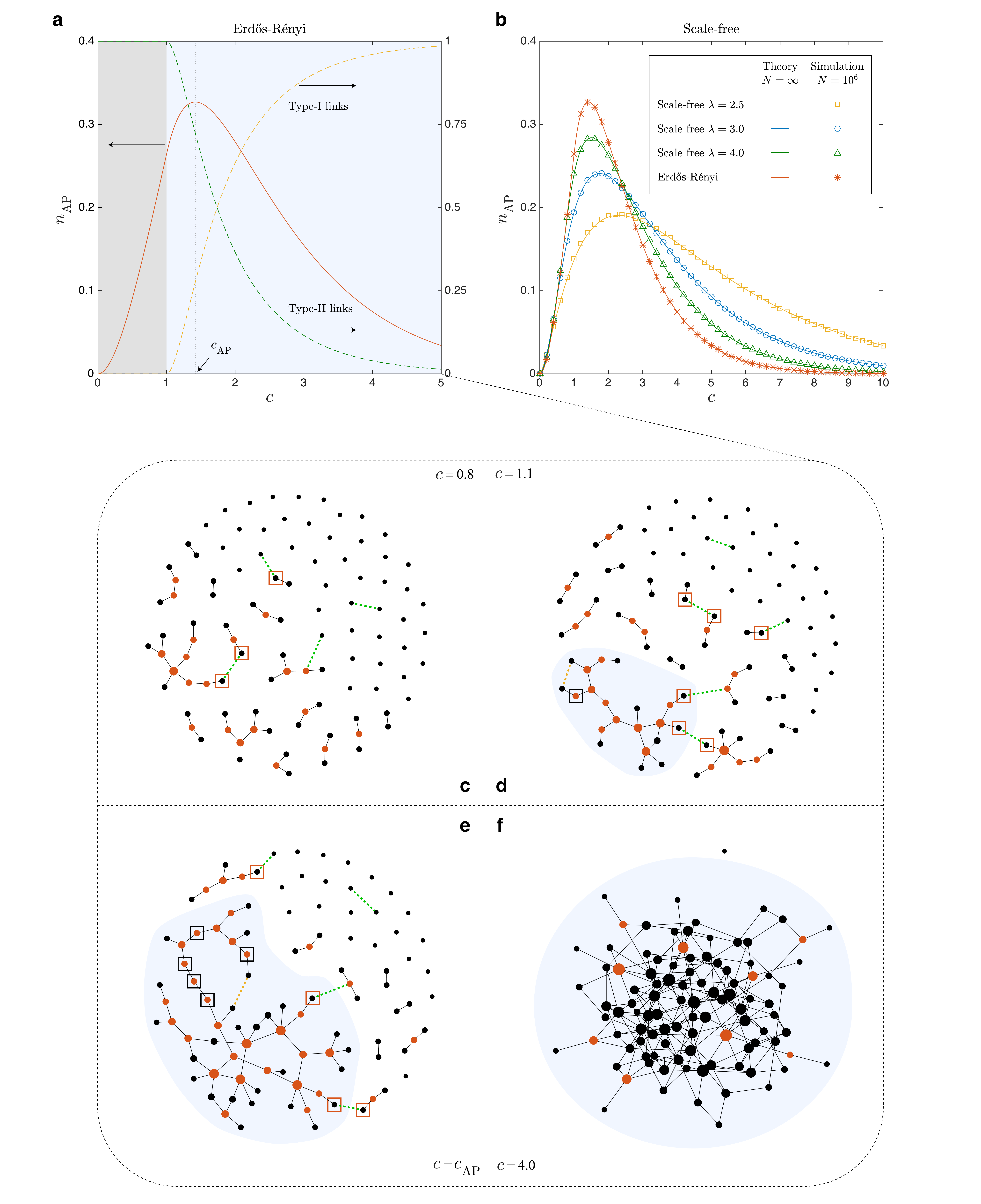}
\caption{}
\end{figure}
\clearpage
\noindent FIG.3. {\bf  \hspace{2mm} Fractions of APs as functions of mean degree in two canonical model networks.} (\textbf{a}) Erd\H os-R\'enyi random networks \cite{erd6s1960evolution}; (\textbf{b}) Scale-free networks with different degree exponents $\lambda$. 
In (\textbf{a}), the fraction of APs is shown in red line. The probabilities of adding type-I (yellow dashed line) and type-II links (green dashed line) are also shown. 
In (\textbf{b}), we use the static model \cite{goh2001universal} to construct SF networks with asymptotically power-law degree distribution $P(k) \sim k^{-\lambda}$. 
Simulation results (symbols) are generated with network size $N=10^6$. Error bars (s.e.m. over 128 realizations) are smaller than the symbols. Lines are our theoretical predictions.
(\textbf{c-f}) Illustrations of APs (red nodes), type-I links (yellow dashed lines), and type-II links (green dashed lines) in ER random networks at different mean degrees. 
Note that adding a single type-II link at most convert two normal nodes to APs (orange boxes), while adding a single type-I link could convert much more APs back to normal nodes (black boxes). This explains why the peak of $n_{\text{AP}}$ emerges even though the probability of adding type-II links is still larger than that of adding type-I links.
The largest connected component is highlighted in light blue in \textbf{d}-\textbf{f}.

\begin{figure}
\centering \includegraphics[width=\textwidth]{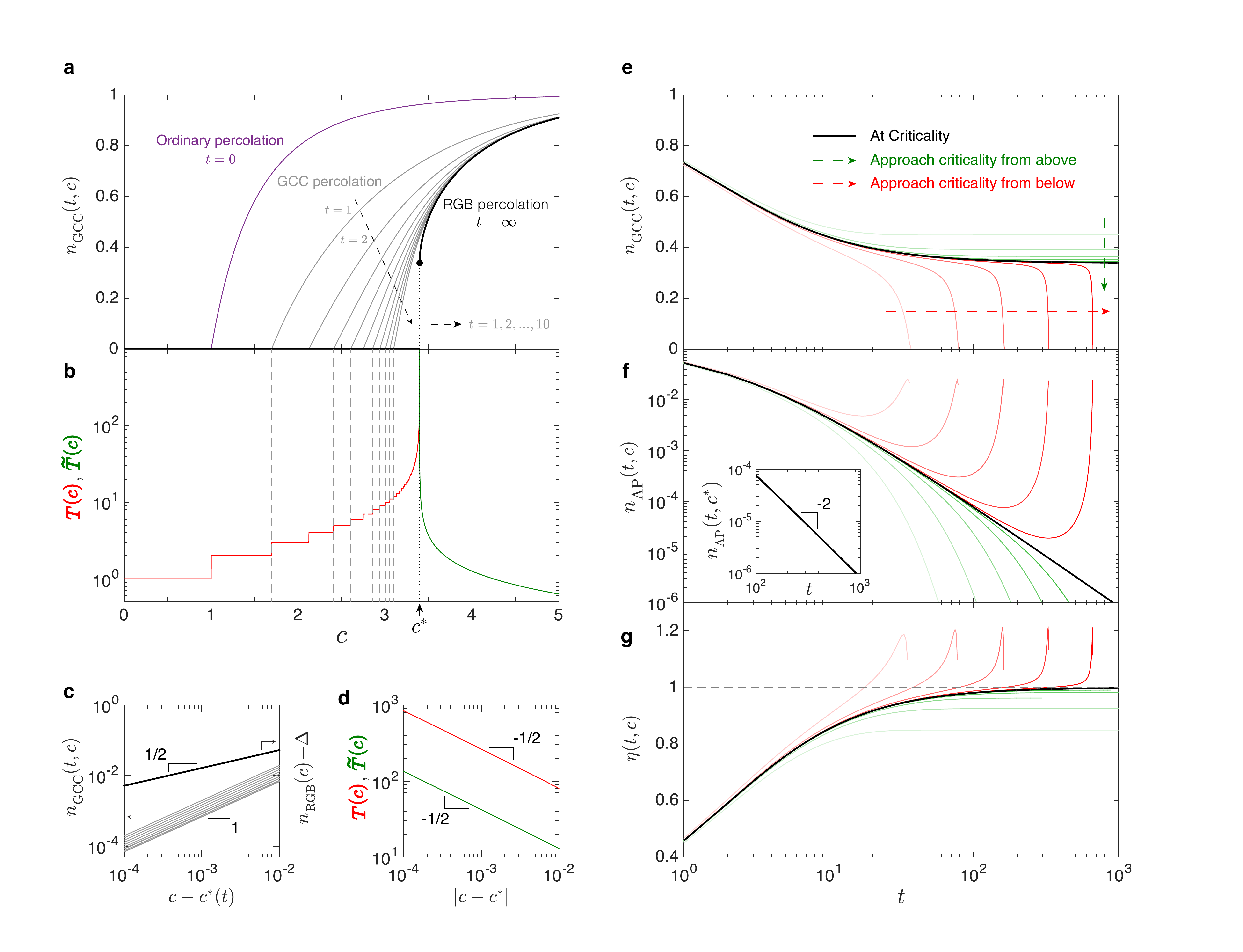}
\caption{{\bf Percolation transitions associated with the greedy articulation points removal process in Erd\H os-R\'enyi random networks.} 
(\textbf{a}) Relative sizes of GCCs after $t$ steps of GAPR, $n_{\text{GCC}}(t,c)$, as functions of the mean degree $c$. Note that $n_{\text{GCC}}(0,c)$ corresponds to the ordinary percolation (thin purple line); $n_{\text{GCC}}(t,c)$ with finite $t$ (only $t=1, 2, ..., 10$ are shown here) corresponds to the GCC percolation (thin grey lines); and $n_{\text{GCC}}(\infty,c)$ corresponds to the RGB percolation (thick black line). 
(\textbf{b}) Total number of the GAPR steps $T(c)$ for $c<c^*$ (red line) and the characteristic number of the GAPR steps $\widetilde{T}(c)$ for $c>c^*$ (green line) as functions of the mean degree $c$.
(\textbf{c}) The critical scaling behavior of $n_{\text{GCC}}(t,c)$ and $n_{\text{RGB}}(c)$ for the GCC and RGB percolation transitions, respectively.
(\textbf{d}) The divergence of $T(c)$ and $\widetilde{T}(c)$ associated with the RGB percolation transition.
Temporal behavior of $n_{\text{GCC}}(t,c)$ (\textbf{e}),  $n_{\text{AP}}(t,c)$ (\textbf{f}), and $\eta(t,c)$ (\textbf{g}) at critical (thick black lines), subcritical (red lines, $c-c^*= -2^4\times10^{-5}, -2^6\times10^{-5}, -2^8\times10^{-5}, -2^{10}\times10^{-5}, -2^{12}\times10^{-5}$, respectively), and supercritical (green lines, $c-c^*= 2^4\times10^{-5}, 2^6\times10^{-5}, 2^8\times10^{-5}, 2^{10}\times10^{-5}, 2^{12}\times10^{-5}$, respectively) regions of the RGB percolation transition. 
At criticality, $n_{\text{AP}}(t,c^*)$ decays in a power-law manner for large $t$ (inset of \textbf{f}).}
\end{figure}
\clearpage

\begin{figure}[h!]
\includegraphics[width=\textwidth]{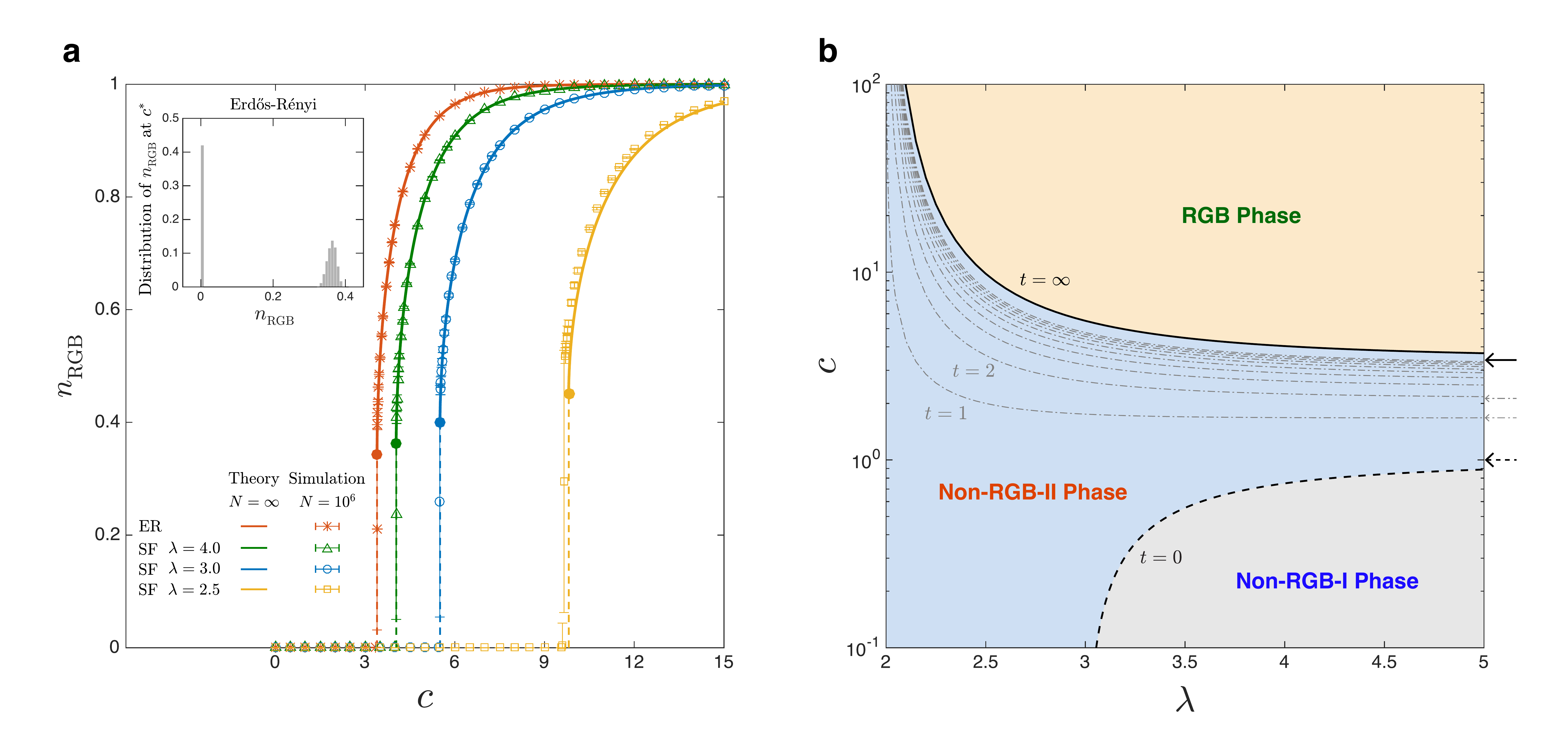}
\caption{{\bf Discontinuous RGB percolation transition and
  the phase diagram.}  
(\textbf{a}) The relative size of the RGB as a function of the mean degree $c$ in the Erd\H os-R\'enyi network (red), and scale-free networks with different degree exponents, $\lambda=$4.0 (green), 3.0 (blue), and 2.5 (yellow), constructed from the static model \cite{goh2001universal}. Lines are our theoretical predictions. 
Simulation results (symbols) are calculated from networks of size $N=10^6$. Error bars (s.e.m. over 128 realizations) are generally smaller than the symbols, except at criticality. 
The deviation of simulation results from our theoretical prediction for $\lambda=2.5$ owes to degree correlations present in the constructed networks, which become prominent as $\lambda \rightarrow 2$ \cite{boguna2004cut}.
Inset displays the distribution of $n_{\text{RGB}}$ generated from 51,200 ER networks of size $N=10^6$ at criticality.
The bimodal distribution of $n_{\text{RGB}}$ indicates that it undergoes a discontinuous jump from nearly zero to a large finite value at the critical point.
(\textbf{b}) Phase diagram associated with the GAPR process in SF networks.
The RGB percolation transition, the GCC percolation transitions (only $t=1, 2, ..., 10$ are presented here), and the ordinary percolation transition are shown in thick solid line, thin dot-dashed lines, and thick dashed line, respectively.
In the limit of large $\lambda$, the phase boundaries for ER networks are recovered 
(highlighted by arrows: $c^{*}=c^{*}(t=\infty)$ (thick solid arrow); $c^{*}(t=1)$ and $c^{*}(t=2)$ (thin dot-dashed arrows); $c_{\text{p}}=c^{*}(t=0)$ (thick dashed arrow)).}
\end{figure}
\newpage

\appendix
\section{Network Destruction}
\subsection{Articulation-point-targeted attack}\label{Sec.SI-I-A}
A common measure of the structural integrity of a network is the size of its giant connected component (GCC).
Articulation points (APs) are natural targets if we aim for immediate damage to the GCC. 
To quickly destruct the GCC, we can target the most destructive AP whose removal will cause the most nodes disconnected from the GCC. This leads to a brute-force strategy of network destruction:
\begin{description}
\item[Step-1] Identify all the APs in the current network, which can be achieved by a linear-time algorithm based on depth-first search~\cite{tarjan1985efficient}. 
\item[Step-2] For each AP, calculate its ``destructivity", i.e. how many nodes will be disconnected from the GCC after its removal.
\item[Step-3] Rank all the APs based on their destructivities, and remove the most destructive one together with the links attaching to it. (In case the most destructive APs are not unique, randomly choose one of them to remove.)
\item[Step-4] Repeat steps 1-3 until 
the network is totally destroyed or left with a residual giant biconnected component (RGB), in which no APs exist.
\end{description}
Hereafter we call this iterative process AP-targeted attack (APTA). 

To demonstrate the efficiency of APTA in reducing the size of the GCC, we compare it with two other network destruction strategies based on the following node centrality measures: 
\begin{itemize}
\item \textbf{Degree}. In this strategy, we iteratively remove the node with the highest degree in the current network \cite{albert2000error, cohen2000resilience}. (In case the highest-degree nodes are not unique, we randomly choose one of them to remove.) Degrees of the remaining nodes will be recalculated after each node removal.
\item \textbf{Collective influence}. The collective influence (CI) of node $i$ is defined as the product of its reduced degree
$(k_i - 1)$ and the total reduced degree of all nodes at distance $l$ from it, i.e. $\text{CI}_l (i) = (k_i - 1) \sum_{j\in \partial \text{Ball}(i,l)} (k_j-1)$,
where $\partial\text{Ball} (i, l)$ is the frontier of a ball of radius $l$ centered at node $i$ (i.e. the set of the $l$-th nearest neighbors of node $i$) \cite{morone2015influence}. In each step, we compute the CI of every node, and remove the node with the largest CI. (In case the highest-CI nodes are not unique, we randomly choose one of them to remove.) It has been shown that, in locally tree-like large networks, the CI-based attack gives the optimal threshold, i.e. the minimal fraction of removed nodes that totally destroys the GCC \cite{morone2015influence}.
Note that, for $l=0$, this method is reduced to the degree-based network destruction. 
\end{itemize}

\begin{figure}
\centering \includegraphics[width=0.8\textwidth]{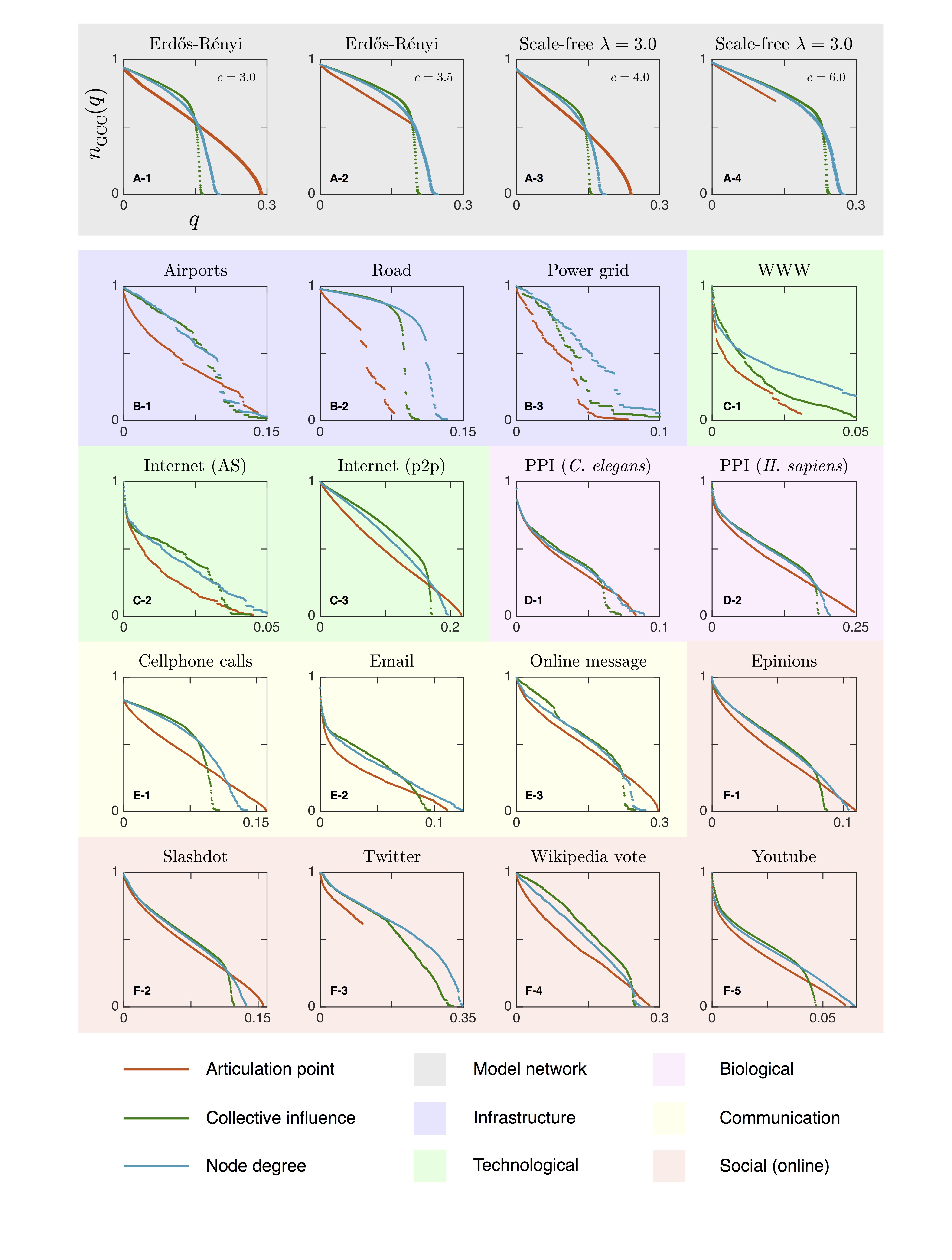}%
\caption{\label{fig.1}Comparison of different network destruction strategies in reducing the size of the giant connected component for both model (\textbf{A}) and real networks (\textbf{B-F}). For model networks, the data points and error bars (defined as s.e.m.) are determined from 32 independent network instances of size $N=10^5$. The real-world networks used in this figure are: openflights (\textbf{B-1}), RoadNet-TX (\textbf{B-2}), PG-WestState (\textbf{B-3}), nd.edu (\textbf{C-1}), oregon2-010331 (\textbf{C-2}), p2p-Gnutella31 (\textbf{C-3}), \textit{Caenorhabditis elegans} (\textbf{D-1}), \textit{Homo sapiens} (\textbf{D-2}), Cellphone (\textbf{E-1}), Email-Enron (\textbf{E-2}), UCIonline (\textbf{E-3}), Epinions (\textbf{F-1}), Slashdot-1 (\textbf{F-2}), Twitter (\textbf{F-3}), WikiVote (\textbf{F-4}), Youtube (\textbf{F-5}) (see section IV for details).}
\end{figure}

We compute the relative size of the GCC, denoted as $n_{\text{GCC}}$,
as a function of the fraction of removed nodes $q$ by using the three
above-mentioned network destruction strategies in two classical model
networks (Fig.~\ref{fig.1}A) and five different types of real-world networks (including
technological, infrastructure, biological, communication, and online
social networks) (Fig.~\ref{fig.1}B-F). Interestingly, we find that, given a limited
``budget" (i.e. a small fraction of nodes to be removed), APTA is the
most efficient strategy in reducing GCC for both model and real
networks (Fig.~\ref{fig.1}). It turns out that even though AP removal maximizes
only the {\it local} damage to the network in every step, and does not
aim for a systemic or global breakdown of the network, for a certain
range of $q$ this is highly efficient in damaging the entire
system. Moreover, in some of the networks, such as the road network (Fig.~\ref{fig.1}B-2), the power grid (Fig.~\ref{fig.1}B-3), and the World
Wide Web (WWW) (Fig.~\ref{fig.1}C-1), APTA is optimal even in the whole
range of $q$. The reason that for those networks APTA leads to smaller
critical $q$ than the state-of-the-art CI-based method is because
these networks are either spatially embedded or/and rich in loops, in
which the CI-based method is not guaranteed to be efficient.  

Note that, even for networks with rather big RGBs (Fig. 1A-2, A-4, F-3), ATPA is still very efficient till the GCC becomes biconnected (or becomes an RGB), where the process of APTA terminates.

\subsection{ Residual giant biconnected component }\label{Sec.SI-I-B}
The RGB can also be obtained by the process of greedy APs removal (GAPR). Performing GAPR on a given network consists of the following steps:
\begin{description}
\item[Step-1] Identify all the APs in the current network by the depth-first search algorithm~\cite{tarjan1985efficient}. 
\item[Step-2] Remove all the APs and the links attaching to them.
\item[Step-3] Repeat steps 1-2 until no APs exist in the network. 
\end{description}

For a general network of finite size, GAPR and APTA
do not necessarily yield exactly the same RGB.
In fact, the RGB
obtained from the former is unique (because the GAPR process is deterministic), while the RGB obtained from the latter could be history-dependent (there may be multiple most destructive APs in each step).

For a given network, we denote the sets of nodes in the two RGBs associated with the two processes as $V_{\text{GAPR}}$ and $V_{\text{APTA}}$, respectively. The overlap between the two RGBs can be measured by the classical Jaccard index:
\begin{equation}\label{eq.1}
\text{overlap}=\frac{\left \vert V_{\text{GAPR}} \bigcap V_{\text{APTA}} \right \vert}{\left \vert V_{\text{GAPR}} \bigcup V_{\text{APTA}}\right  \vert}.
\end{equation}

\begin{figure}
\centering \includegraphics[scale=0.25]{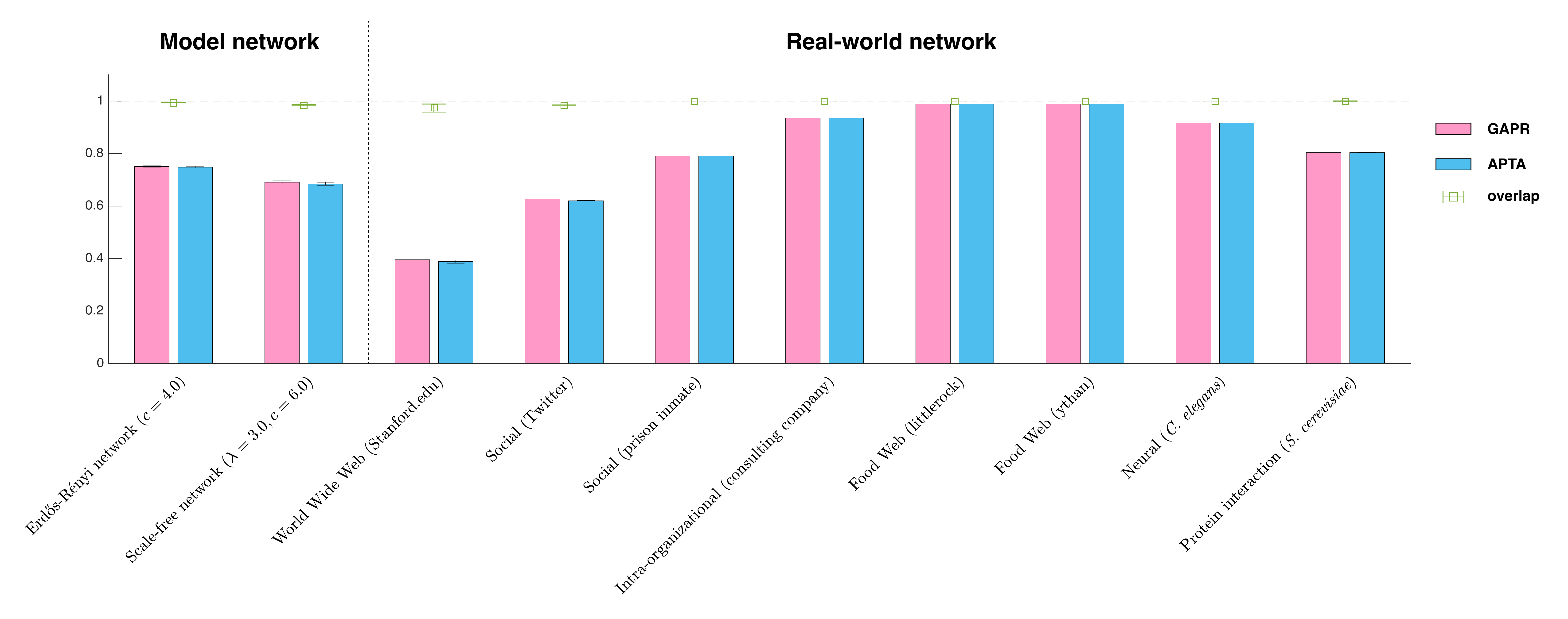}%
\caption{\label{fig.2}The relative sizes of the two RGBs associated with the GAPR (pink bars) and the APTA (blue bars) processes, and the overlap between the two RGBs (green squares) in model networks and real-world networks. For each type of model network, the results (with error bars defined as s.e.m.) are obtained from 32 independent network instances with $N=10^5$. For each real-world network, the results (with error bars defined as s.e.m.) are averaged over 32 independent realizations of APTA processes.}
\end{figure}

In Fig. \ref{fig.2}, we show the relative sizes of the two RGBs, as well as the overlap between them.
We find that the two RGBs significantly overlap for both model networks and real-world networks.

\section{Theoretical Framework} \label{Sec.SI-II}

\subsection{Tree ansatz}
In our theoretical framework, we study the GAPR process in large uncorrelated random networks with arbitrary degree
distribution $P(k)$ and finite mean degree $c=\sum_{k}kP(k)$. In the thermodynamic limit, these networks are locally tree-like and do not contain loops of finite length. This tree ansatz directly leads to three important network properties \cite{cohen2000resilience, callaway2000network, newman2001random, dorogovtsev2008critical, mezard2009information}: 
\begin{enumerate}
\item All of the finite connected components (FCCs) are trees. Loops (of infinite length) only exist in the GCC; 
\item There can be only one GCC in a network;
\item For any node in a network, its neighbors become disconnected or independent from each other if the node and its attaching links are removed.
\end{enumerate}

We emphasize that, even though our theoretical framework is based on the tree ansatz, its predictions are fairly precise for finite and loopy networks, provided that the density of loops is not excessively large (see section III for numerical simulations).

\subsection{Discrete-time dynamics of GAPR}
The deterministic GAPR process is naturally associated with a temporal order. At $t=0$, we have the original network. At $t=T$, i.e. the last time step of GAPR, no AP exists in the network, and the GAPR process terminates. At each time step $t$, we denote the fraction of APs and the relative size of the GCC as $n_{\text{AP}}(c,t)$ and $n_{\text{GCC}}(c,t)$, respectively, where $c$ is the mean degree of the original network. The RGB is nothing but the GCC at the last time step, whose relative size is denoted as $n_{\text{RGB}}(c)=n_{\text{GCC}}(c,T)$.

At each time step $t$ during the GAPR process in a network $\mathcal{G}$, we classify the remaining nodes into the following three categories or states: 
\begin{enumerate}
\item $\alpha_t$-nodes: nodes that belong to FCCs; 
\item $\beta_t$-nodes: nodes that are APs; and belong to the GCC; 
\item $\gamma_t$-nodes: nodes that are not APs; and belong to the GCC.
\end{enumerate}
Note that the notations $\beta$ and $\gamma$ here have totally different meanings from the critical exponents mentioned in the main text.

According to the tree ansatz, the state of a randomly chosen node $i$ can be determined by the states of its neighbors in $\mathcal{G} \backslash i$, i.e. the induced subgraph of $\mathcal{G}$ with node $i$ and all its links removed. Therefore, at each time step $t$, we need to know the probability that, following a randomly chosen link to one of its end nodes,  this node belongs to any of the above categories after this link is removed. These probabilities are denoted as $\alpha_t$, $\beta_t$, and $\gamma_t$, respectively. To be precise, hereafter when we consider the state of a neighbor of a given node $i$, we mean the state of the neighbor in the induced subgraph $\mathcal{G} \backslash i$. Also, when we mention the state of an end node of a chosen link $l$, we mean the state of the node in $\mathcal{G} \backslash l$, i.e. the induced subgraph of $\mathcal{G}$ with link $l$ removed. According to the definitions of $\beta_t$ and $\gamma_t$, the probability that an end node of a randomly chosen link belongs to the GCC at time step $t$ is just $\beta_t+\gamma_t$.

\begin{figure}
\centering \includegraphics[scale=0.6]{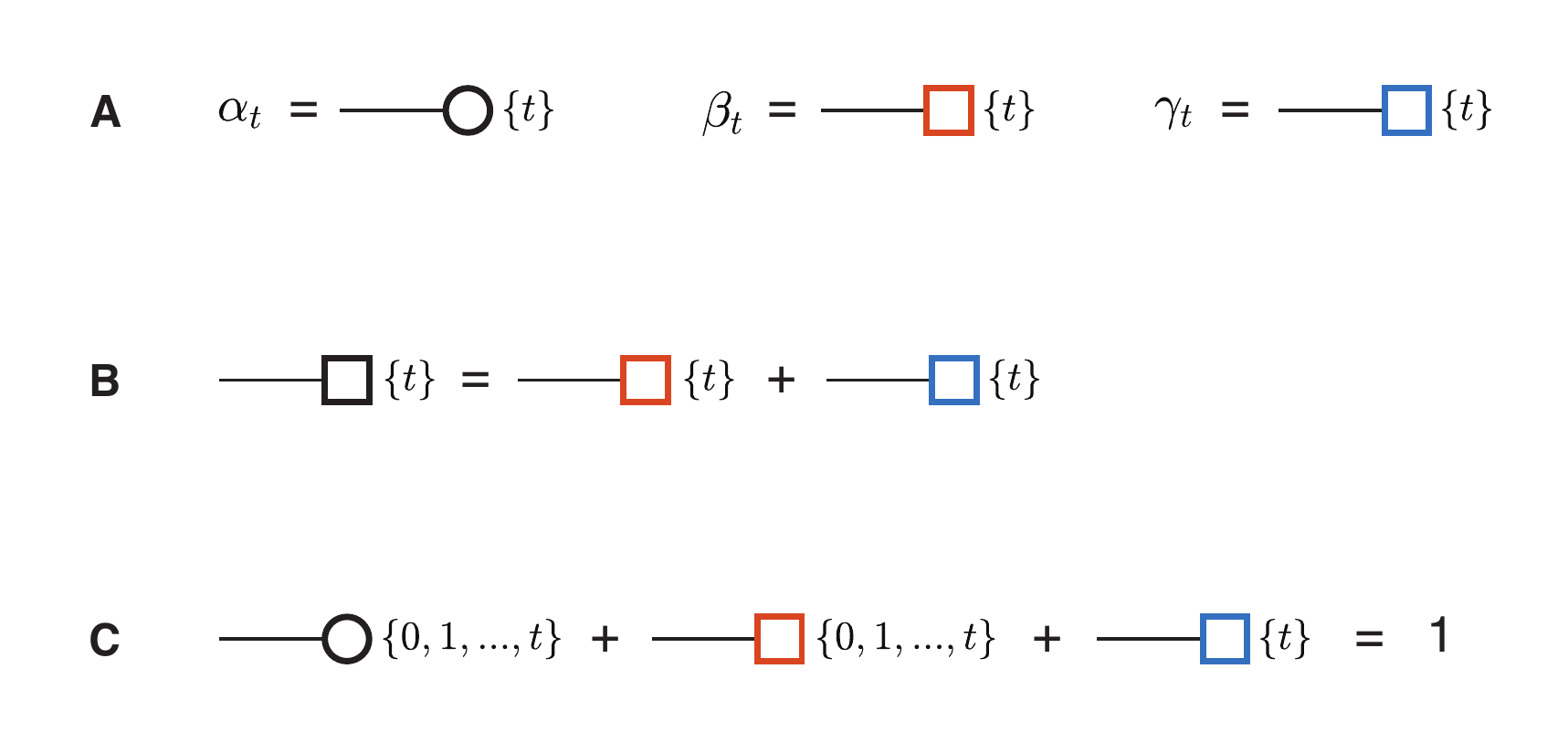}%
\caption{\label{fig.3}(A) Diagrammatic notations for the probabilities that, following a randomly chosen link to one of its end nodes, we arrive at $\alpha_t$-node, $\beta_t$-node, and $\gamma_t$-node, respectively, at time step $t$. (B) The probability that an end node of a randomly chosen link belongs to the GCC at time step $t$ is given by $\beta_t+\gamma_t$. (C) Diagrammatic representation of Eq. \ref{eq.2} for the normalization condition of the three sets of probabilities $\{\alpha_0, \alpha_1,...\}$, $\{\beta_0, \beta_1,...\}$, and $\{\gamma_0, \gamma_1,...\}$.}
\end{figure}

\begin{figure}
\centering \includegraphics[scale=0.6]{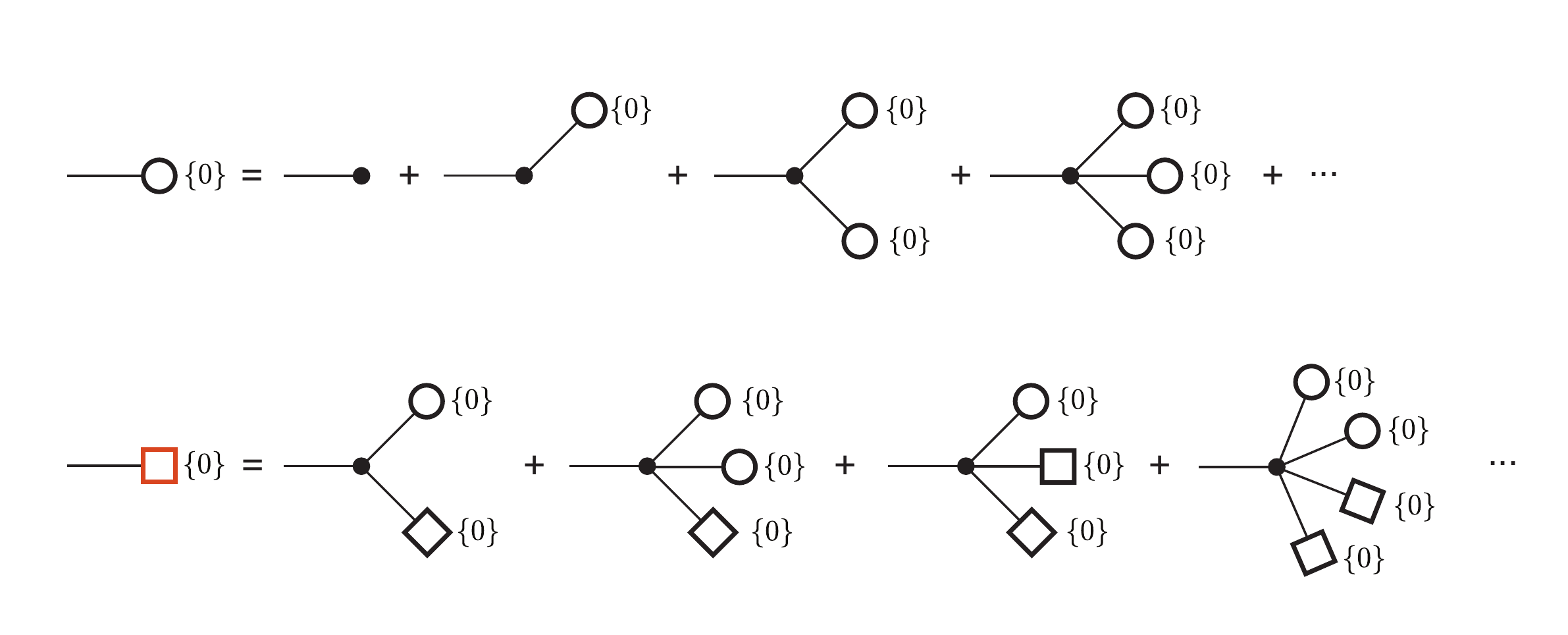}%
\caption{\label{fig.4}Diagrammatic representations of Eqs. \ref{eq.3a} and \ref{eq.3b}.}
\end{figure}

The GAPR process can be fully characterized by the three sets of probabilities $\{\alpha_0, \alpha_1,...\}$, $\{\beta_0, \beta_1,...\}$, and $\{\gamma_0, \gamma_1,...\}$. 
Note that every node must belong to one of the three categories, which means the three sets of probabilities are not independent from each other. Specifically, at time step $t$, the probability $\gamma_t$ can be derived by the other two sets of probabilities through the following normalization condition:
\begin{equation}\label{eq.2}
\sum_{\tau=0}^t  \alpha_{\tau}  + \sum_{\tau=0}^t  \beta_{\tau}   + \gamma_t = 1.
\end{equation}
Figure~\ref{fig.3} shows the diagrammatic representations of these probabilities; the probability that an end node of a randomly chosen link belongs to the GCC at time step $t$; and the normalization equation \ref{eq.2}.
Thanks to Eq. \ref{eq.2}, hereafter all the equations will be written in terms of $\alpha_t$ and $\beta_t$ only. We can calculate $\{\alpha_0, \alpha_1,...\}$ and $\{\beta_0, \beta_1,...\}$ in an iterative way. At first, we consider the initial time step $t=0$. The self-consistent equations for $\alpha_0$ and $\beta_0$ are given by
\begin{subequations}\label{eq.3}
\begin{equation}\label{eq.3a}
\alpha_0=\sum_{k=1}^{\infty}Q(k)\left( \alpha_0  \right)^{k-1}
\end{equation}
\begin{equation}\label{eq.3b}
\beta_0=\sum_{k=3}^{\infty}Q(k)\left[ 1- \left( 1-\alpha_0 \right)^{k-1} - (\alpha_0 ) ^ {k-1} \right],
\end{equation}
\end{subequations}
where 
\begin{equation}\label{eq.4}
Q(k)=kP(k)/c
\end{equation}
is the degree distribution of the nodes that we arrive at by following a randomly chosen link (a.k.a. the excess degree distribution) \cite{albert2002statistical, newman2003structure}. We derive the above equations based on the following observations:
\begin{itemize}
\item $\alpha_0$-node: its neighbors can only be $\alpha_0$-nodes;
\item $\beta_0$-node: since it is an AP node, at least one of its neighbors is an $\alpha_0$-node. Moreover, since it belongs to the GCC, at lease one of its neighbors is not an $\alpha_0$-node.
\end{itemize}
Note that, due to the tree ansatz, removal of an AP in the GCC can only cause a finite number of nodes disconnected from the GCC, and there exists no such an AP that its removal breaks down the GCC into two infinitely large components. 
The diagrammatic representations of Eqs.~\ref{eq.3a} and~\ref{eq.3b} are shown in Fig.~\ref{fig.4}.

\begin{figure}
\centering\includegraphics[scale=0.6]{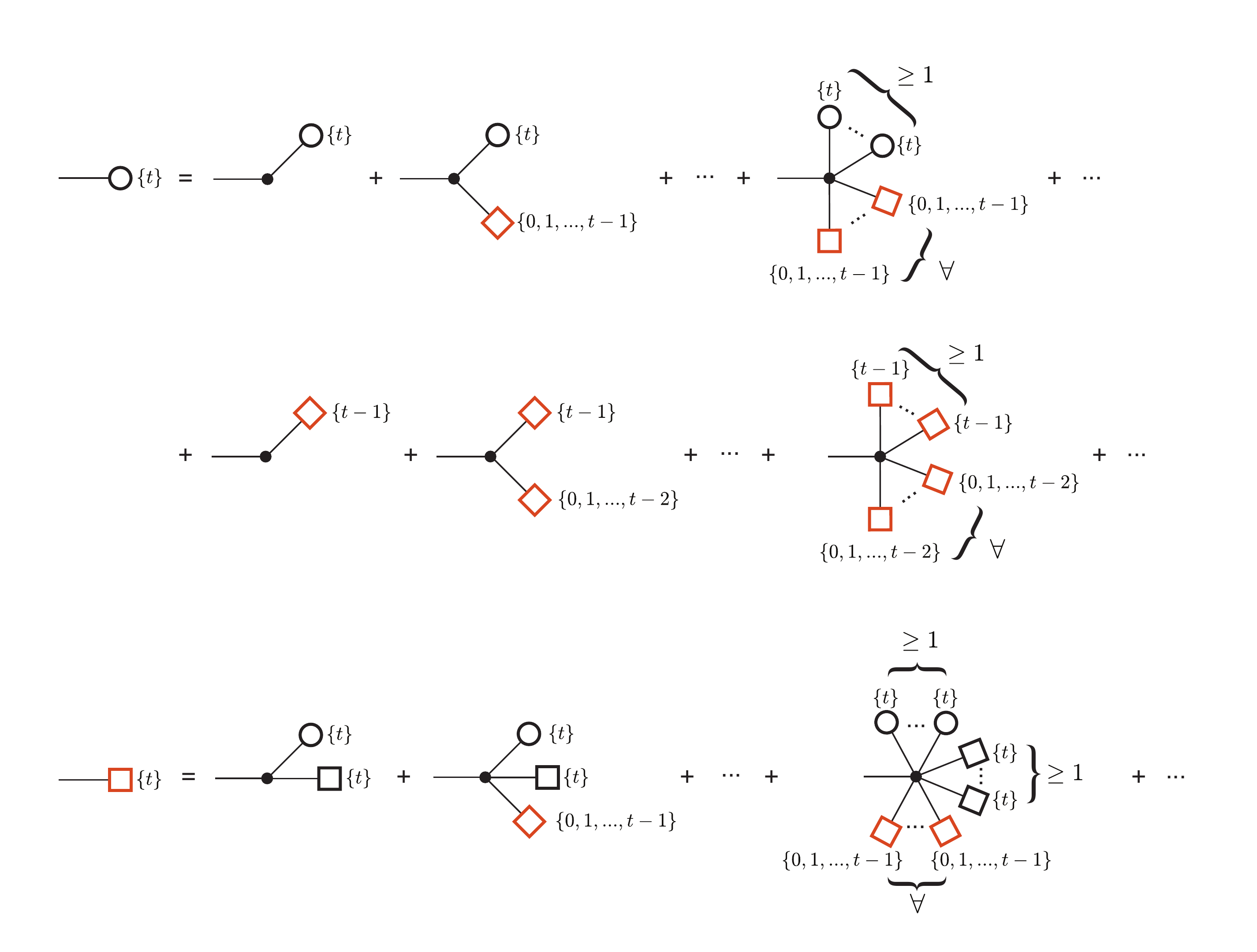}%
\caption{\label{fig.5}Diagrammatic representations of Eqs. \ref{eq.5a} and \ref{eq.5b}.}
\end{figure}

For the $t$-th GAPR time step ($t>0$), we can compute $\alpha_t$ and $\beta_t$ as follows:
\begin{subequations}
\label{eq.5}
\begin{equation}
\label{eq.5a}
\alpha_t=\sum_{k=1}^{\infty}Q(k)\left( \alpha_t+ \sum_{\tau=0}^{t-1} \beta_{\tau} \right)^{k-1}-\sum_{k=1}^{\infty}Q(k)\left( \sum_{\tau=0}^{t-2} \beta_{\tau}  \right)^{k-1}
\end{equation}
\begin{eqnarray}
\label{eq.5b}
\beta_t =&&\sum_{k=3}^{\infty}Q(k)\sum_{s=1}^{k-2}\binom{k-1}{s}\left(1- \sum_{\tau=0}^{t} \alpha_{\tau} - \sum_{\tau=0}^{t-1} \beta_{\tau} \right)^s \nonumber \\
&& \times \sum_{r=1}^{k-1-s}\binom{k-1-s}{r}\left( \alpha_t \right)^{r} \left( \sum_{\tau=0}^{t-1} \beta_{\tau} \right)^{k-1-s-r}.
\end{eqnarray}
\end{subequations}
The derivations of Eqs. \ref{eq.5a} and \ref{eq.5b} are based on the following observations:
\begin{itemize}
\item $\alpha_t$-node: First of all, its neighbors can only be $\alpha_t$-nodes or $\beta_{\tau}$-nodes with $\tau<t$ (because if one of its neighbors is $\alpha_{\tau}$-node with $\tau<t$, this node will be an AP before time step $t$ and hence would have already been removed; if one of its neighbor is $\beta_t$-node, this node will belong to the GCC at time step $t$). Second, its neighbors can not be simultaneously all $\beta_{\tau}$-nodes with $\tau<t-1$. Otherwise this node will be a leaf node before time step $t-1$, and will become an isolated node before the $t$-th time step. In this case, we can not reach this node through a randomly chosen link at time step $t$.
\item $\beta_t$-node: First of all, its neighbors can not be $\alpha_{\tau}$-nodes with $\tau<t$, otherwise this node would have been removed before time step $t$. Second, since it is an AP node, at least one of its neighbors is an $\alpha_t$-node. Finally, since this node belongs to the GCC, at least one of its neighbors is neither $\alpha_t$-node nor $\beta_{\tau}$-node with $\tau<t$.
\end{itemize}
The diagrammatic representations of Eqs. \ref{eq.5a} and \ref{eq.5b} are shown in Fig.~\ref{fig.5}.

Combining Eqs. \ref{eq.3} and \ref{eq.5}, we obtain a complete set of discrete-time dynamic equations for the GAPR process:
\begin{subequations}\label{eq.6}
\begin{equation}\label{eq.6a}
\alpha_0=G_1(\alpha_0)
\end{equation}
\begin{equation}\label{eq.6b}
\alpha_{t>0}=G_1\left(\alpha_t+\sum_{\tau=0}^{t-1}\beta_\tau\right)-G_1\left(\sum_{\tau=0}^{t-2}\beta_\tau\right)
\end{equation}
\begin{equation}\label{eq.6c}
\beta_t=G_1\left(1-\sum_{\tau=0}^{t-1}\alpha_\tau\right)-G_1\left(1-\sum_{\tau=0}^{t}\alpha_\tau\right)-G_1\left(\alpha_t+\sum_{\tau=0}^{t-1}\beta_\tau\right)+G_1\left(\sum_{\tau=0}^{t-1}\beta_\tau\right),
\end{equation}
\end{subequations}
where 
\begin{equation}\label{eq.7}
G_1(x)=\sum_k Q(k) x^{k-1}
\end{equation}
is the generating function of the excess degree distribution $Q(k)$. 

By solving the above self-consistent equations, we obtain $\{\alpha_0, \alpha_1,...\}$ and $\{\beta_0, \beta_1,...\}$, which govern the whole process of GAPR. With these two sets of probabilities, we can compute any quantities of interest, such as the total number of GAPR steps, the fraction of APs, the relative size of the GCC and the RGB, and so on.

\subsection{Total number of GAPR steps}
It should be noted from Eq. \ref{eq.6} that $\alpha_t$ and $\beta_t$ depend not only on the previous time step $t-1$, but also on the entire history of GAPR, from the 0-th to the $(t-1)$-th time steps. Hence to calculate $T$, the total number of GAPR steps, we have to solve the equations from the initial step to the $T$-th step when the GAPR process stops.  
Since the GAPR process terminates when there is no APs left in the network, $T$ can be determined by requiring
\begin{subequations}
\begin{equation}\label{eq.8a}
\alpha_T=0
\end{equation}
\begin{equation}\label{eq.8b}
\beta_T=0.
\end{equation}
\end{subequations}
Note that, according to Eq. \ref{eq.6b}, $\beta_{T-1}$ is also zero, otherwise $\alpha_T$ will have a non-zero solution.

\subsection{Fraction of articulation points}
\begin{figure}
\centering\includegraphics[scale=0.6]{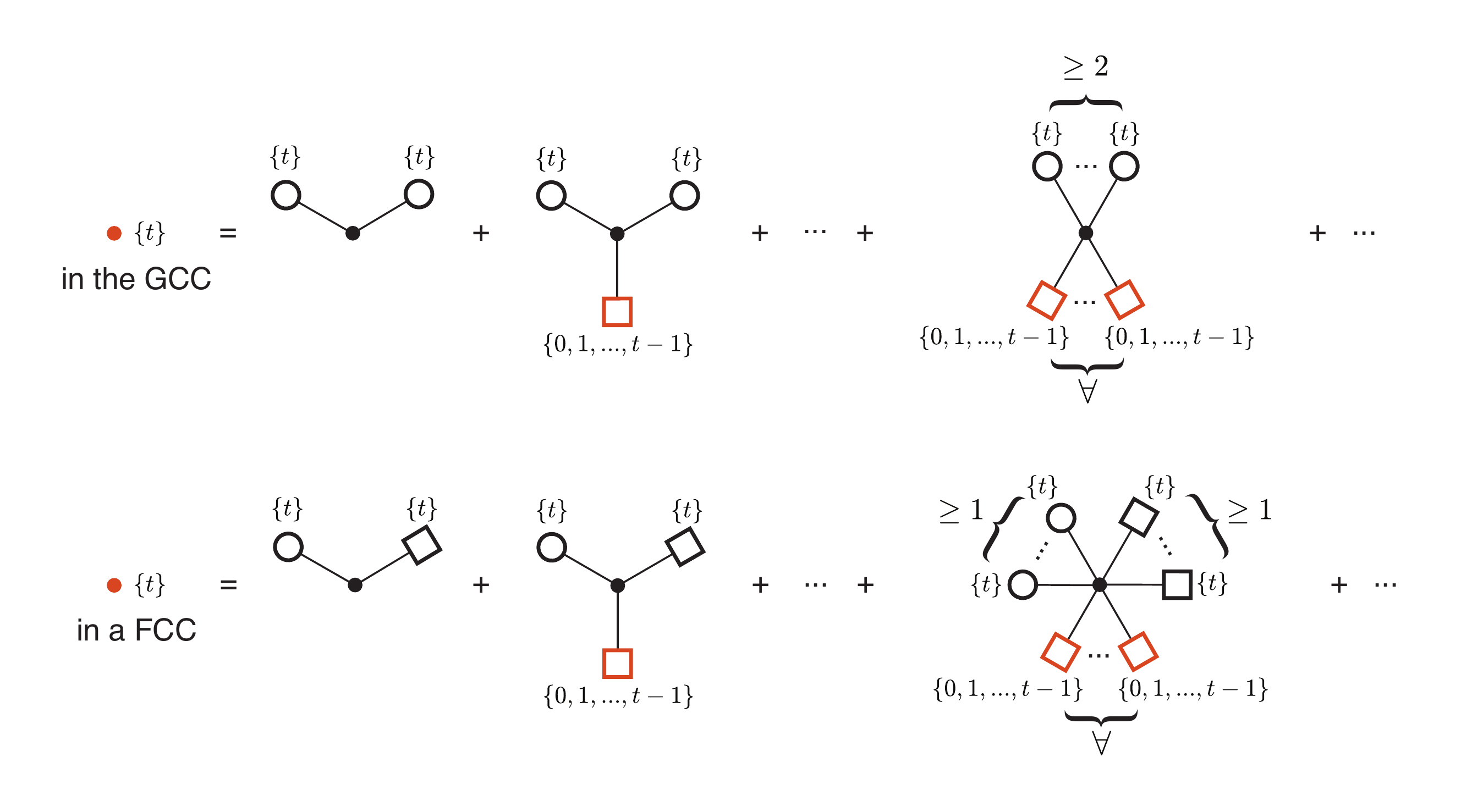}%
\caption{\label{fig.6}Diagrammatic representations of Eqs. \ref{eq.9a} and \ref{eq.9b}.}
\end{figure}

Consider the fraction of APs at time step $t$ during the GAPR process, $n_\text{AP}(t,c)$, which also represents the probability that a randomly chosen node is an AP at $t$-th step.
We can calculate $n_{\text{AP}}(t,c)$ as the sum of the following two parts: 
\begin{enumerate}
\item $n_{\text{AP}}^{\text{GCC}}(t,c)$: fraction of APs that belong to the GCC; 
\item $n_{\text{AP}}^{\text{FCC}}(t,c)$: fraction of APs that belong to FCCs. 
\end{enumerate}
With similar theoretical considerations as deriving equations for $\alpha_t$ and $\beta_t$, we have
\begin{subequations}
\begin{equation}\label{eq.9a}
n_{\text{AP}}^{\text{FCC}}(t,c)=\sum_{k=2}^{\infty}P(k) \sum_{s=2}^{k} \binom{k}{s} \left( \alpha_t \right)^s \left( \sum_{\tau=0}^{t-1} \beta_{\tau} \right)^{k-s}
\end{equation}
\begin{eqnarray}\label{eq.9b}
n_{\text{AP}}^{\text{GCC}}(t,c) =&&\sum_{k=2}^{\infty}P(k)\sum_{s=1}^{k-1}\binom{k}{s}\left(1- \sum_{\tau=0}^{t} \alpha_{\tau} - \sum_{\tau=0}^{t-1} \beta_{\tau} \right)^s \nonumber \\
&& \times \sum_{r=1}^{k-s}\binom{k-s}{r}\left( \sum_{\tau=0}^{t} \alpha_{\tau} \right)^{r} \left( \sum_{\tau=0}^{t-1} \beta_{\tau} \right)^{k-s-r}.
\end{eqnarray}
\end{subequations}
These two equations can be understood as follows:
\begin{itemize}
\item AP in FCCs, $n_{\text{AP}}^{\text{FCC}}(t,c)$:  First of all, since the node belongs to an FCC, its neighbor can only be $\alpha_t$-nodes or $\beta_{\tau}$-nodes with $\tau<t$. Second, since the node is an AP, at least two of its neighbors are $\alpha_t$-nodes. Otherwise, it would be a leaf node or an isolated node at time step $t$.  
\item AP in the GCC, $n_{\text{AP}}^{\text{GCC}}(t,c)$: First of all, its neighbors can not be $\alpha_{\tau}$-nodes with $\tau<t$, otherwise this node would have already been removed before time step $t$. Second, since it is an AP node, at least one of its neighbors is an $\alpha_t$-node. Finally, since this node belongs to the GCC, at least one of its neighbors is neither $\alpha_t$-node nor $\beta_{\tau}$-node with $\tau<t$.
\end{itemize}
The diagrammatic representations of Eqs. \ref{eq.9a} and \ref{eq.9b} are shown in Fig.~\ref{fig.6}.

Combing Eqs. \ref{eq.9a} and \ref{eq.9b}, we obtain the fraction of APs at time step $t$ as
\begin{equation}\label{eq.10}
n_\text{AP}(t,c)=n_{\text{AP}}^{\text{FCC}}(t,c)+n_{\text{AP}}^{\text{GCC}}(t,c)=G_0\left(1-\sum_{\tau=0}^{t-1}\alpha_\tau\right)-G_0\left(1-\sum_{\tau=0}^{t}\alpha_\tau\right) - c \alpha_{t} G_1\left( \sum_{\tau=0}^{t-1}\beta_{\tau} \right),
\end{equation}
where 
\begin{equation}\label{eq.11}
G_0(x)=\sum_k P(k) x^k
\end{equation}
is the generating function of the degree distribution $P(k)$.

The fraction of APs in the original network, $n_{\text{AP}}(c)$, can be obtained by substituting $t=0$ into Eq.~\ref{eq.10} the above equation, yielding
\begin{equation}\label{eq.12}
n_\text{AP}(c)=n_\text{AP}(0,c)=1-G_0\left(1-\alpha_0\right) - c \alpha_{0} G_1\left( 0 \right).
\end{equation}

\subsection{The giant connected component}
\begin{figure}
\centering\includegraphics[scale=0.6]{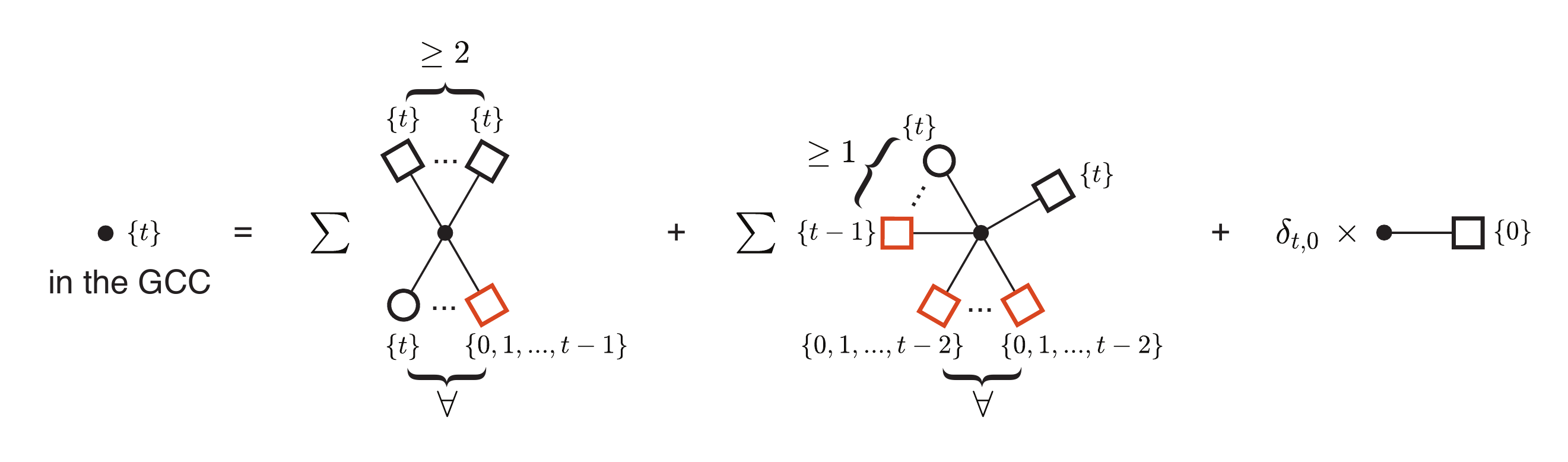}%
\caption{\label{fig.7}Diagrammatic representation of Eq. \ref{eq.13}.}
\end{figure}

Another quantity of interest is the relative size of the GCC, $n_{\text{GCC}}(t,c)$, after performing a given $t$ steps of GAPR in a network. We can compute $n_\text{GCC}(t,c)$ as follows: 
\begin{eqnarray}\label{eq.13}
n_{\text{GCC}}(t)=&&\sum_{k=2}^{\infty}P(k)\sum_{s=2}^{k}\binom{k}{s}\left( \alpha_{t} + \sum_{\tau=0}^{t-1}\beta_{\tau}\right)^{k-s}\left(1-\sum_{\tau=0}^{t}\alpha_{\tau} - \sum_{\tau=0}^{t-1}\beta_{\tau} \right)^s \nonumber \\
&& + \sum_{k=2}^{\infty}P(k)\binom{k}{1} \left( 1-\sum_{\tau=0}^{t}\alpha_{\tau} - \sum_{\tau=0}^{t-1}\beta_{\tau} \right) \sum_{r=1}^{k-1} \binom{k-1}{r} \left( \alpha_{t} + \beta_{t-1}  \right)^r \left( \sum_{\tau=0}^{t-2}\beta_{\tau} \right)^{k-1-r} \nonumber \\
&& + \delta_{t,0} P(1) \left( 1-\alpha_{0} \right),
\end{eqnarray}
where the Kronecker delta function $\delta_{jk}=1$ if $j=k$, and 0 otherwise. This equation can be understood as follows: 
\begin{itemize}
\item For any node in the GCC at time step $t$, at least one of its neighbors belongs to the GCC at current time step. Moreover, none of its neighbors is $\alpha_{\tau}$-nodes with $\tau<t$, otherwise this node would have already been removed as an AP before time step $t$.
\item The first term on the right hand side of Eq. \ref{eq.13} accounts for the nodes with at least two neighbors belonging to the GCC.
\item The second term accounts for the nodes with only one neighbor in the GCC. It should be noted that, for each of these nodes, at least one of its neighbors is $\alpha_t$-nodes or $\beta_{t-1}$-nodes. Otherwise, this node would be a leaf node before time step $t$, and thus will become an isolated node at time step $t$.
\item We only consider nodes with degree larger than 2 in the first two terms. This is because any leaf nodes will become isolated ones right after the first step of GAPR. If we include the ordinary percolation as the 0-th time step in our GAPR process, leaf nodes can also belong to the GCC at the 0-th step, which is the third term.
\end{itemize}
The diagrammatic representation of Eq. \ref{eq.13} is shown in Fig.~\ref{fig.7}. 
After some algebra, we can rewrite $n_{\text{GGC}}(t,c)$ in terms of generating functions:
\begin{equation}\label{eq.14}
n_{\text{GGC}}(t,c)=G_0\left(1-\sum_{\tau=0}^{t-1}\alpha_{\tau}\right)-G_0\left( \alpha_{t} + \sum_{\tau=0}^{t-1}\beta_{\tau} \right)- \left( 1- \delta_{\tau,0} \right) c\left(1-\sum_{\tau=0}^{t}\alpha_{\tau} - \sum_{\tau=0}^{t-1}\beta_{\tau} \right) G_1\left(\sum_{\tau=0}^{t-2}\beta_{\tau} \right).
\end{equation}
Note that the relative size of the GCC for ordinary percolation can be recovered by substituting $t=0$ into the above equation, yielding 
\begin{equation}\label{eq.15}
n_{\text{GCC}}(0)=1-G_0\left( \alpha_0 \right).
\end{equation}
\subsection{The residual giant biconnected component}
\begin{figure}
\centering\includegraphics[scale=0.6]{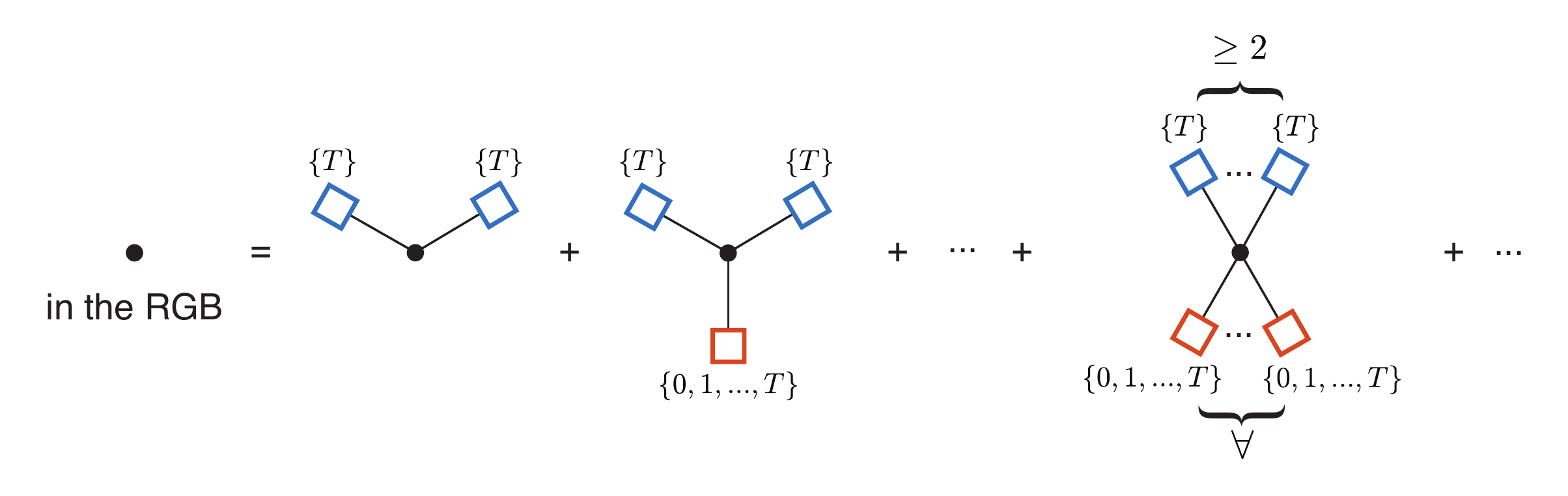}%
\caption{\label{fig.8}Diagrammatic representation of Eq. \ref{eq.16}.}
\end{figure}

If we keep performing GAPR in a network until there is no AP left, depending on the initial structure of the network, it could be totally destructed or left with an RGB. The relative size of the RGB can be calculated as
\begin{equation}\label{eq.16}
n_{\text{RGB}}(c)=\sum_{k=2}^{\infty} P(k) \sum_{s=2}^{k} \binom{k}{s} \left( 1- \sum_{\tau=0}^{T}\alpha_{\tau} - \sum_{\tau=0}^{T}\beta_{\tau} \right) \sum_{\tau=0}^{T}\beta_{\tau},
\end{equation}
which is based on the following observations:
\begin{itemize}
\item For any node in the RGB, its neighbors can not be $\alpha_t$-nodes at any $t$. Otherwise, this node is removable as an AP.
\item Since the RGB is biconnected, each of its nodes must be connected to at least two other RGB nodes. The probability that an end node of a randomly chosen link belongs to the RGB is $1- \sum_{\tau=0}^{T}\alpha_{\tau} - \sum_{\tau=0}^{T}\beta_{\tau}$.
\end{itemize}
The diagrammatic representation of Eq. \ref{eq.16} is shown in Fig.~\ref{fig.8}. After some algebra, Eq. \ref{eq.16} can be rewritten in terms of generating functions as:
\begin{equation}\label{eq.17}
n_{\text{RGB}}(c)=G_0\left(1-\sum_{\tau=0}^{T}\alpha_{\tau}\right)-G_0\left(\sum_{\tau=0}^{T}\beta_{\tau}\right)-c \left(1-\sum_{\tau=0}^{T}\alpha_{\tau}-\sum_{\tau=0}^{T}\beta_{\tau}\right) G_1\left(\sum_{\tau=0}^{T}\beta_{\tau}\right).
\end{equation}

Note that the RGB is nothing but the GCC at the last time step $T$, we can also obtain the above equation by substituting $t=T$ and also $\alpha_T=0$, $\beta_T=0$,  $\beta_{T-1}=0$ into the Eq. \ref{eq.14} of $n_{\text{GCC}}(t,c)$.

\subsection{Canonical model networks }
The developed theoretical framework works for complex networks with any prescribed degree distribution. In this work, we consider two canonical model networks: Erd\H os-R\'enyi (ER) networks and scale-free (SF) networks, which have specific degree distributions.

\subsubsection{Erd\H os-R\'enyi network}
For Erd\"os-R\'enyi random network, the degree distribution $P(k)$ in the thermodynamic limit follows the Poisson distribution, i.e.
\begin{equation}
\label{eq.18}
P(k)=\frac{e^{-c}c^k}{k!},
\end{equation}
where $c=\sum_{k=0}^{\infty} kP(k)$ is the mean degree. The generating functions are given by
\begin{equation}
\label{eq.19}
G_0(x)=G_1(x)=e^{c(x-1)}.
\end{equation}

Substituting the above equation into Eqs. \ref{eq.6a}, \ref{eq.6b}, and \ref{eq.6c}, we obtain the dynamic equations of the GAPR process on Erd\H os-R\'enyi network:
\begin{subequations}
\begin{equation}\label{eq.20a}
\alpha_0=\exp \left[ c\left( \alpha_0-1 \right) \right]
\end{equation}
\begin{equation}\label{eq.20b}
\alpha_{t>0}=\exp \left[ c\left(\alpha_t+\sum_{\tau=0}^{t-1}\beta_\tau-1\right) \right] - \exp \left[ c \left( \sum_{\tau=0}^{t-2}\beta_\tau-1 \right) \right]
\end{equation}
\begin{equation}\label{eq.20c}
\beta_t=\exp \left(-c \sum_{\tau=0}^{t-1}\alpha_\tau\right)- \exp \left(-c\sum_{\tau=0}^{t}\alpha_\tau\right)-\exp \left[ c\left(\alpha_t+\sum_{\tau=0}^{t-1}\beta_\tau-1\right)\right]+\exp \left[ c\left(\sum_{\tau=0}^{t-1}\beta_\tau-1\right)\right].
\end{equation}
\end{subequations}
Similarly, the fraction of APs is given by
\begin{equation}\label{eq.21}
n_\text{AP}(t,c)=\exp \left(-c\sum_{\tau=0}^{t-1}\alpha_\tau\right)-\exp \left(-c\sum_{\tau=0}^{t}\alpha_\tau\right) - c \alpha_{t} \exp \left[ c\left( \sum_{\tau=0}^{t-1}\beta_{\tau}-1 \right)\right].
\end{equation}
Hence the fraction of APs in the original network at $t=0$ is given by
\begin{equation}\label{eq.22}
n_\text{AP}(c)=n_\text{AP}(0,c)=1-\exp \left( -c\alpha_0 \right) -c\alpha_0 \exp \left( -c \right).
\end{equation}
The relative size of the GCC at the $t$-th time step is given by
 \begin{eqnarray}\label{eq.23}
n_{\text{GCC}}(t)=&& \exp \left(-c\sum_{\tau=0}^{t-1}\alpha_{\tau}\right)- \exp \left[ c\left( \alpha_{t} + \sum_{\tau=0}^{t-1}\beta_{\tau}-1 \right) \right] \nonumber \\
&& - \left( 1- \delta_{\tau,0} \right) c\left(1-\sum_{\tau=0}^{t}\alpha_{\tau} - \sum_{\tau=0}^{t-1}\beta_{\tau} \right) \exp \left[ c\left(\sum_{\tau=0}^{t-2}\beta_{\tau}-1 \right) \right].
\end{eqnarray}
The relative size of the RGB is
\begin{equation}\label{eq.24}
n_{\text{RGB}}(c)=\exp \left(-c\sum_{\tau=0}^{T}\alpha_{\tau}\right)-\exp \left[ c\left(\sum_{\tau=0}^{T}\beta_{\tau} - 1\right) \right] -c \left(1-\sum_{\tau=0}^{T}\alpha_{\tau}-\sum_{\tau=0}^{T}\beta_{\tau}\right) \exp \left[ c\left(\sum_{\tau=0}^{T}\beta_{\tau} -1 \right)\right].
\end{equation}

\subsubsection{Scale-free network}
We use the static model to generate asymptotically SF networks with tunable network size $N$, mean degree $c$, and degree distribution exponent $\lambda>2$ \cite{goh2001universal}.

This model can be described as follows: 
\begin{description}
\item[Step-1] We start with $N$ isolated nodes, which are labeled from 1 to $N$. Each node is assigned a probability $p_i \sim i^{-a}$, where $a=\frac{1}{\lambda-1}$.
\item[Step-2] We independently pick up two nodes according to their assigned probabilities, and add a link between these two nodes if they have not been connected before. Self-links and double-links are forbidden.
\item[Step-3] Repeat step-2 until $M=cN/2$ links have been added into the network.
\end{description}

In the thermodynamic limit, the degree distribution of the static model can be analytically derived \cite{catanzaro2005analytic, lee2006intrinsic}:
\begin{equation}\label{eq.25}
P(k)=\frac{[c(1-a)]^k}{a k!}\int_1^{\infty} dx \exp[-c(1-a)x] x^{k-1-1/a}.
\end{equation}
The generating function of the degree distribution $P(k)$ is given by
\begin{eqnarray}\label{eq.26}
G_0(x)=&& \sum_{k=0}^{\infty} \frac{[c(1-a)]^k}{a k!}\int_1^{\infty} dy \exp[-c(1-a)y] y^{k-1-1/a} x^k \nonumber\\
=&& \frac{1}{a} \int_1^{\infty} dy \exp[-c(1-a)y] y^{-1-1/a} \sum_{k=0}^{\infty} \frac{[c(1-a)xy]^k}{k!} \nonumber \\
=&& \frac{1}{a} \int_1^{\infty} dy \exp[-c(1-a)(1-x)y] y^{-1-1/a},
\end{eqnarray}
where $e^{c(1-a)xy}=\sum_{k=0}^{\infty} \frac{[c(1-a)xy]^k}{k!}$ is used in the last step. Similarly, the generation function of the excess degree distribution $Q(k)$ is derived as 
\begin{equation}\label{eq.27}
G_1(x)= \frac{G'_0(x)}{G'_0(1)}= \frac{1-a}{a} \int_1^{\infty} dy \exp[-c(1-a)(1-x)y] y^{-1/a}. 
\end{equation}
By defining
\begin{equation}\label{eq.28}
E_n(x)=\int_1^{\infty} e^{-xy}y^{-n} dy,
\end{equation}
$G_0(x)$ and $G_1(x)$ can be rewritten as
\begin{subequations}
\begin{equation}\label{eq.29a}
G_0(x)  =  \frac{1}{a} E_{1+\frac{1}{a}} [c(1-a)(1-x)]
\end{equation}
\begin{equation}\label{eq.29b}
G_1(x)  =  \frac{1-a}{a} E_{\frac{1}{a}} [c(1-a)(1-x)] 
\end{equation}
\end{subequations}
where the fact $\frac{\partial E_n(x)}{\partial x}=-E_{n-1}(x)$ has been used.

Substituting the above equations into Eqs. \ref{eq.6a}, \ref{eq.6b}, and \ref{eq.6c}, we obtain the dynamic equations of the GAPR process for the static model:
\begin{subequations}
\begin{equation}\label{eq.30a}
\alpha_0=\frac{1-a}{a} E_{\frac{1}{a}} \left[ c \left( 1-a \right) \left(1-\alpha_0 \right) \right] 
\end{equation}
\begin{equation}\label{eq.30b}
\alpha_{t>0}= \frac{1-a}{a} \left\{ E_{\frac{1}{a}} \left[ c \left( 1-a \right) \left(1-\alpha_t - \sum_{\tau=0}^{t-1}\beta_\tau \right) \right] - E_{\frac{1}{a}} \left[ c \left( 1-a \right) \left(1- \sum_{\tau=0}^{t-2}\beta_\tau \right) \right] \right\}
\end{equation}
\begin{eqnarray}\label{eq.30c}
\beta_t=&& \frac{1-a}{a} \Bigg\{
E_{\frac{1}{a}} \left[ c \left( 1-a \right) \sum_{\tau=0}^{t-1}\alpha_\tau \right]
- E_{\frac{1}{a}} \left[ c \left( 1-a \right) \sum_{\tau=0}^{t}\alpha_\tau \right] \nonumber \\
&& - E_{\frac{1}{a}} \left[ c \left( 1-a \right) \left( 1-\alpha_t - \sum_{\tau=0}^{t-1}\beta_\tau \right) \right]
- E_{\frac{1}{a}} \left[ c \left( 1-a \right) \left( 1-\sum_{\tau=0}^{t-1}\beta_\tau \right) \right]
\Bigg\}.
\end{eqnarray}
\end{subequations}
Similarly, the fraction of APs is given by
\begin{eqnarray}\label{eq.31}
n_\text{AP}(t,c)=&& \frac{1}{a} \left\{
E_{1+\frac{1}{a}} \left[ c \left( 1-a \right) \sum_{\tau=0}^{t-1}\alpha_\tau \right]
-E_{1+\frac{1}{a}} \left[ c \left( 1-a \right) \sum_{\tau=0}^{t}\alpha_\tau \right] \right\} \nonumber \\
&& - \frac{c\alpha_t(1-a)}{a} E_{\frac{1}{a}} \left[ c \left( 1-a \right) \left( 1- \sum_{\tau=0}^{t-1}\beta_{\tau} \right) \right].
\end{eqnarray}
Hence the fraction of APs in the original network at $t=0$ is
\begin{equation}\label{eq.32}
n_\text{AP}(c)=n_\text{AP}(0,c)=1- \frac{1}{a} E_{1+\frac{1}{a}} [c(1-a)\alpha_0]-  \frac{c\alpha_0(1-a)}{a} E_{\frac{1}{a}} [c(1-a)].
\end{equation}
The relative size of the GCC at the $t$-th time step is given by
 \begin{eqnarray}\label{eq.33}
n_{\text{GCC}}(t)=&& \frac{1}{a} \left\{
E_{1+\frac{1}{a}} \left[ c \left( 1-a \right) \sum_{\tau=0}^{t-1}\alpha_{\tau} \right]
- E_{1+\frac{1}{a}} \left[ c \left( 1-a \right) \left( 1- \alpha_{t} - \sum_{\tau=0}^{t-1}\beta_{\tau} \right) \right] \right\} \nonumber \\
&& -\left( 1- \delta_{\tau,0} \right) \frac{c(1-a)}{a} \left(1-\sum_{\tau=0}^{t}\alpha_{\tau} - \sum_{\tau=0}^{t-1}\beta_{\tau} \right) E_{\frac{1}{a}} \left[ c \left( 1-a \right) \left(1- \sum_{\tau=0}^{t-2}\beta_{\tau} \right) \right].
\end{eqnarray}
The relative size of the RGB is
\begin{eqnarray}\label{eq.34}
n_{\text{RGB}}(c)=&& \frac{1}{a} \left\{
E_{1+\frac{1}{a}} \left[ c \left( 1-a \right) \sum_{\tau=0}^{T}\alpha_{\tau} \right]
- E_{1+\frac{1}{a}} \left[ c \left( 1-a \right) \left( 1- \sum_{\tau=0}^{T} \beta_{\tau} \right) \right] \right\} \nonumber \\
&& - \frac{c(1-a)}{a} \left(1-\sum_{\tau=0}^{T}\alpha_{\tau}-\sum_{\tau=0}^{T}\beta_{\tau} \right)
E_{\frac{1}{a}} \left[ c \left( 1-a \right) \left(1- \sum_{\tau=0}^{T}\beta_{\tau} \right) \right].
\end{eqnarray}

\clearpage

\section{Numerical Simulations}\label{Sec.SI-III}
\begin{figure}[H]
 \includegraphics[scale=0.35]{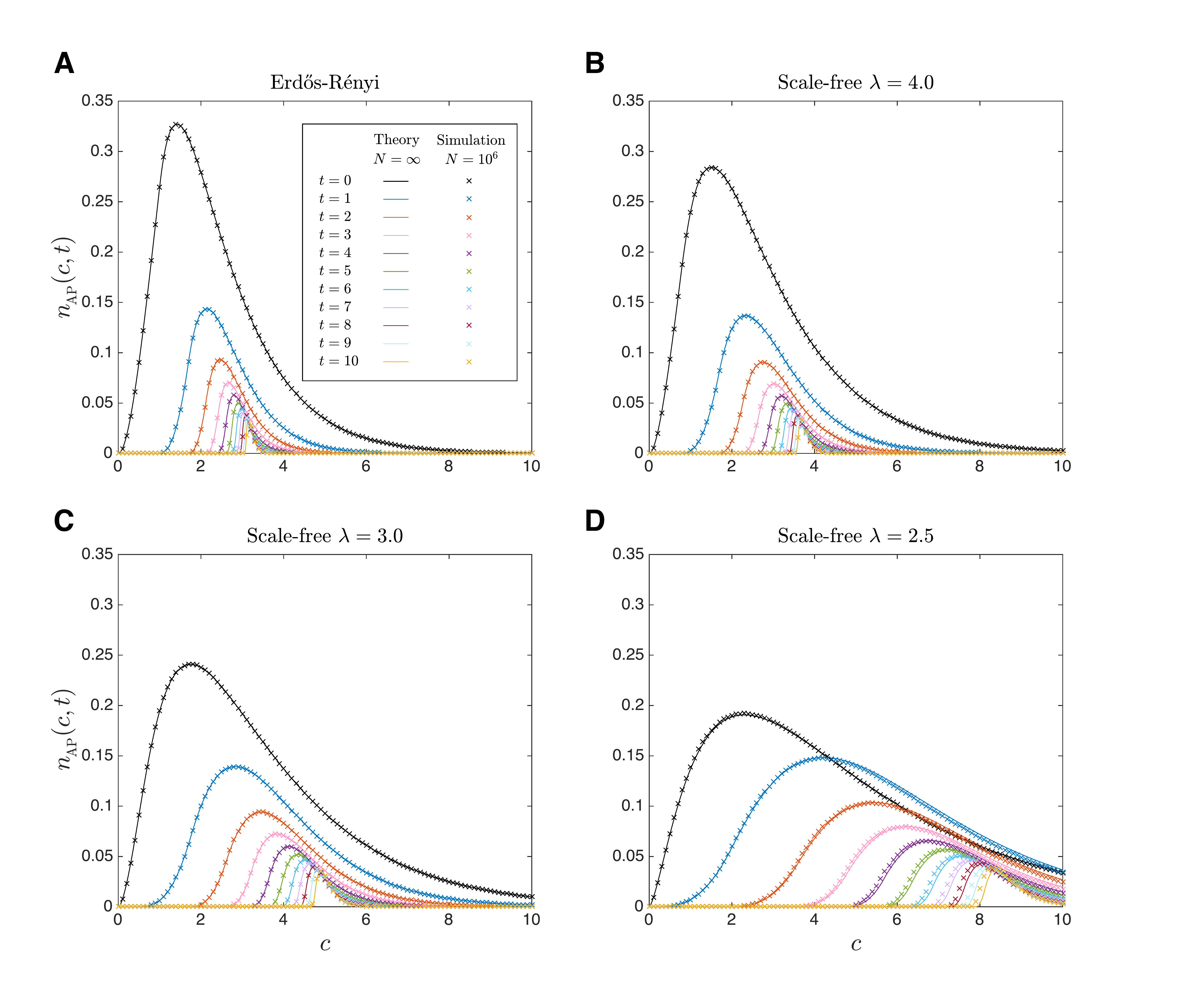}%
 \caption{\label{fig.9}
The fraction of APs, $n_{\text{AP}}(c,t)$, as a function of mean degree $c$ at different time steps $t$ in the GAPR process, for ER networks (A) and SF networks with different degree exponents $\lambda=4.0, 3.0, 2.5$ (B-D). Symbols are numerical results calculated from the GAPR
procedure on finite discrete networks constructed with ER model (A) and the static model (B-D) with $N=10^6$. 
The data points and error bars (defined as s.e.m.) are determined from 128 independent network realizations.
Lines are analytical results for infinitely large system ($N \to \infty$).
 }
 \end{figure}

Figure~\ref{fig.9} shows the simulation results of the fraction of APs, $n_{\text{AP}}(c,t)$, as a function of mean degree $c$ at different time steps for ER networks and SF networks (constructed from the ER model and the static model) with different degree exponents $\lambda$. For comparison, analytical results for infinitely large networks are also shown (in lines). 

\begin{figure}[H]
 \includegraphics[scale=0.45]{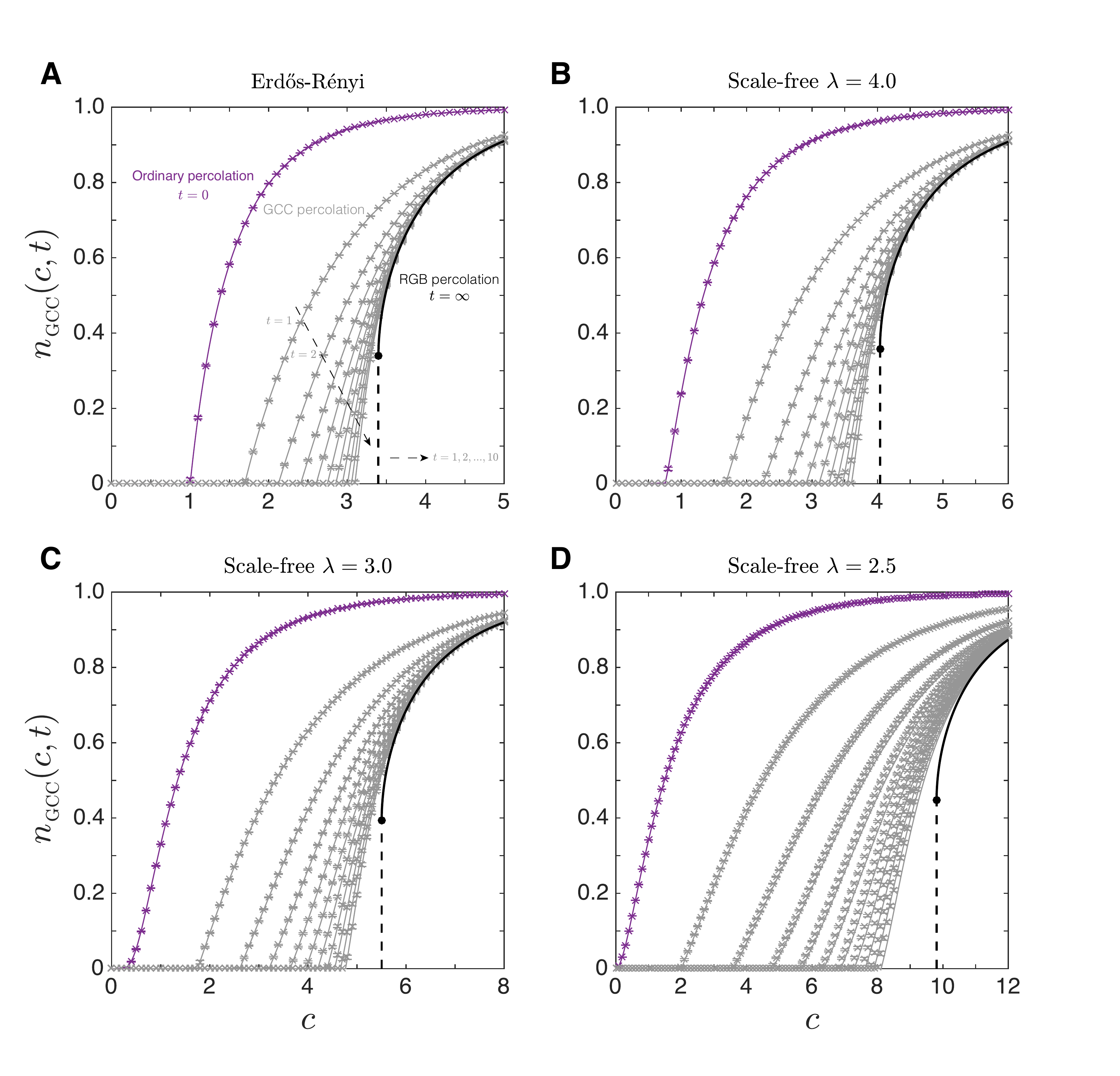}%
 \caption{\label{fig.10}The relative size of the GCC, $n_{\text{GCC}}(c,t)$, as a function of mean degree $c$ at different time steps $t$ in the GAPR process, for ER networks (A) and SF networks with different degree exponents $\lambda=4.0, 3.0, 2.5$ (B-D). Symbols are numerical results calculated by performing a given steps of GAPR
procedure on finite discrete networks constructed with ER model (A) and the static model (B-D) with $N=10^6$. 
The data points and error bars (defined as s.e.m.) are determined from 128 independent network realizations.
Thin lines are analytical results for $n_{\text{GCC}}(c,t)$ with finite $t$, and thick lines are analytical results for the relative size of the RGB, which is $n_{\text{GCC}}(c,\infty)$. 
 }
 \end{figure}

The comparison between simulation results and analytical predictions of the relative size of the GCC, $n_{\text{GCC}}(c,t)$, is shown in Fig.~\ref{fig.10}. Note that, in Figs.~\ref{fig.9} and~\ref{fig.10}, the deviation of the simulation results from the theoretical prediction for SF networks with exponent $\lambda=2.5$ owes to degree correlations in the constructed scale-free networks, which become prominent as $\lambda \rightarrow 2$ \cite{boguna2004cut}.

\begin{figure}[H]
 \includegraphics[scale=0.45]{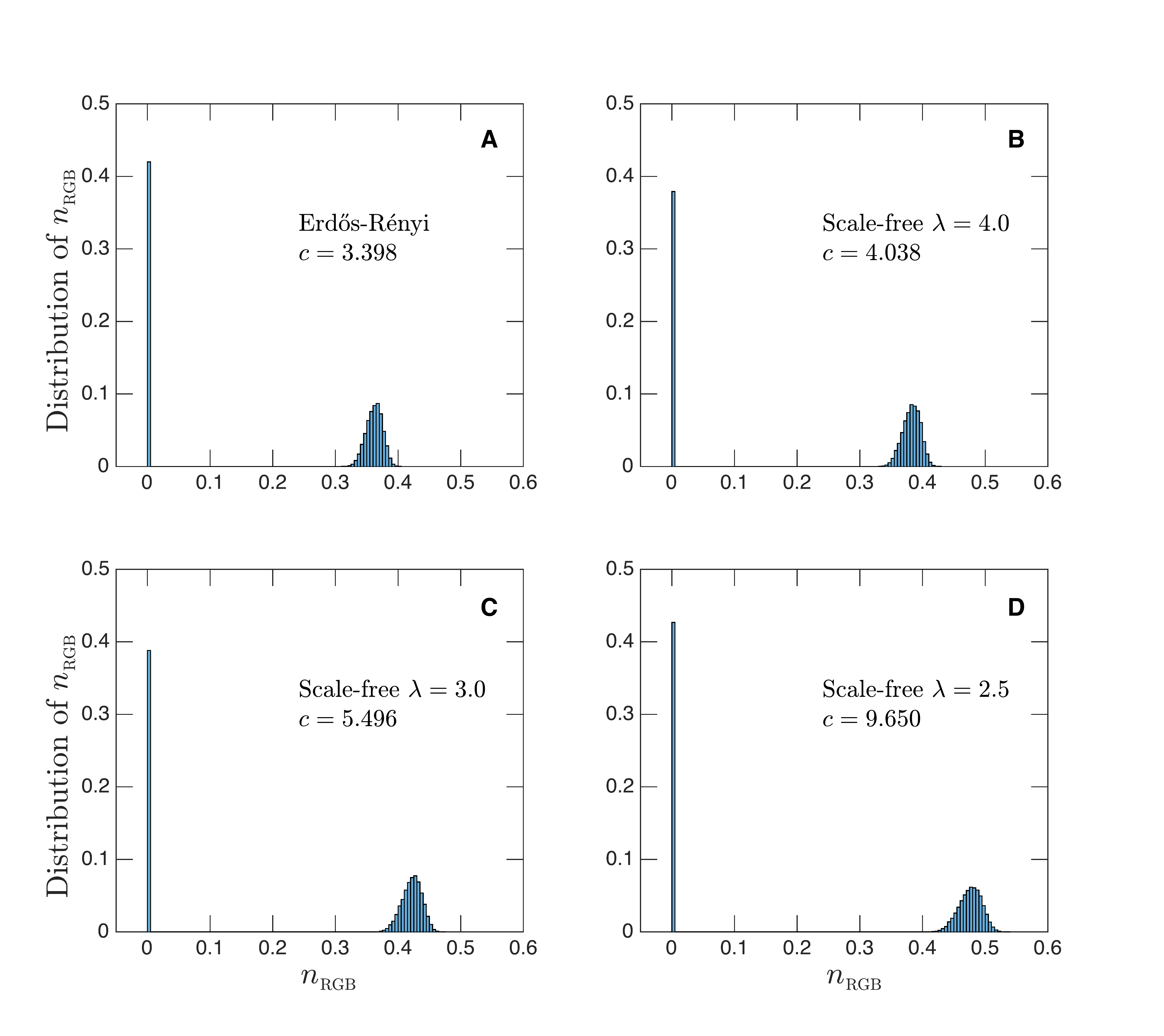}%
 \caption{\label{fig.11}
 The distribution of the relative size of the RGB, $n_{\text{RGB}}$, at criticality for both ER network (A) and SF networks with different exponents $\lambda=4.0, 3.0, 2.5$ (B-D). Each distribution is generated by performing GAPR process on 12,800 different finite discrete networks of size $N=10^6$. 
 }
 \end{figure}

In Fig.~\ref{fig.11}, we show the simulation results of the distribution of the relative size of the RGB, $n_{\text{RGB}}$, at criticality for both ER network and SF networks with different degree exponents. The bimodal distribution is another evidence that $n_{\text{RGB}}$ undergoes a discontinuous jump from zero to a large finite value at the critical point \cite{dorogovtsev2008critical}.
 
\clearpage
 
 \clearpage

 \clearpage

\newpage
\section{Network Datasets} \label{Sec.SI-IV}
All the real-world networks analyzed in this work are listed and briefly described in the following tables. For each network, we show its type, name and reference; number of nodes ($N$) and edges ($L$); and a brief description (unless it is clear from the network type).

\begin{table}[H]
\caption{Air traffic networks}
\begin{ruledtabular}
\begin{tabular}{lrrl}
Name & $N$ & $L$  & description \\\hline
USairport500 \cite{colizza2007reaction}&	500&	5,960&Top 500 busiest commercial airports in US\\
USairport-2010 \cite{opsahl2011anchorage}&	1,574&	28,236&US airport network in 2010\\
 openflights \cite{opsahl2011anchorage}&	2,939&	30,501&non-US-based airport network\\
\end{tabular}
\end{ruledtabular}
\end{table}

\begin{table}[H]
\caption{Road networks}
\begin{ruledtabular}
\begin{tabular}{lrrl}
Name & $N$ & $L$  & description \\\hline
RoadNet-CA \cite{leskovec2014snap}&	1,965,206&	5,533,214&California road network\\
RoadNet-PA \cite{leskovec2014snap}&	1,088,092&	3,083,796&Pennsylvania road network\\
RoadNet-TX \cite{leskovec2014snap}&	1,379,917&	3,843,320&Texas road network\\
\end{tabular}
\end{ruledtabular}
\end{table}

\begin{table}[H]
\caption{Power grids (PG)}
\begin{ruledtabular}
\begin{tabular}{lrrl}
Name & $N$ & $L$  & description \\\hline
PG-Texas \cite{bianconi2008local}& 4,889 & 5,855 & Power grid in Texas \\
PG-WestState \cite{watts1998collective}& 4,941 & 6,594 &High-voltage power grid in the Western States in US\\
\end{tabular}
\end{ruledtabular}
\end{table}

\begin{table}
\caption{Internet: Autonomous Systems (AS) \cite{leskovec2014snap}}
\begin{ruledtabular}
\begin{tabular}{lrrl}
Name & $N$ & $L$  & description \\\hline
as20000102 &6,474&	13,895& AS graph from January 02 2000\\
oregon1-010331&	10,670&	22,003&AS peering information inferred from Oregon route-views (I)\\
oregon1-010407&	10,729&	22,000& Same as above (at different time)\\
oregon1-010414  &	10,790&	22,470& Same as above (at different time)\\
oregon1-010421  &	10,859&	22,748& Same as above (at different time)\\
oregon1-010428  &	10,886&	22,494& Same as above (at different time)\\
oregon1-010505  &	10,943&	22,608& Same as above (at different time)\\
oregon1-010512  &	11,011&	22,678& Same as above (at different time)\\
oregon1-010519  &	11,051&	22,725& Same as above (at different time)\\
oregon1-010526  &	11,174&	23,410& Same as above (at different time)\\
oregon2-010331  &	10,900&	31,181& AS peering information inferred from Oregon route-views (II)\\
oregon2-010407  &	10,981&	30,856& Same as above (at different time)\\
oregon2-010414  &	11,019&	31,762& Same as above (at different time)\\
oregon2-010421  &	11,080&	31,539& Same as above (at different time)\\
oregon2-010428  &	11,113&	31,435& Same as above (at different time)\\
oregon2-010505  &	11,157&	30,944& Same as above (at different time)\\
oregon2-010512  &	11,260&	31,304& Same as above (at different time)\\
oregon2-010519  &	11,375&	32,288& Same as above (at different time)\\
oregon2-010526  &	11,461&	32,731& Same as above (at different time)\\
\end{tabular}
\end{ruledtabular}
\end{table}

\begin{table}
\caption{Internet: peer-to-peer (p2p) file sharing networks \cite{leskovec2014snap}}
\begin{ruledtabular}
\begin{tabular}{lrrl}
Name & $N$ & $L$  & description \\\hline
p2p-Gnutella04 &10,876&	39,994&Gnutella p2p file sharing network\\
p2p-Gnutella05 &	8,846&	31,839& Same as above (at different time)\\
p2p-Gnutella06 &	8,717&	31,525& Same as above (at different time)\\
p2p-Gnutella08 &	6,301&	20,777& Same as above (at different time)\\
p2p-Gnutella09 &	8,114&	26,013& Same as above (at different time)\\
p2p-Gnutella24 &	26,518&	65,369& Same as above (at different time)\\
p2p-Gnutella25 &	22,687&	54,705& Same as above (at different time)\\
p2p-Gnutella30 &	36,682&	88,328& Same as above (at different time)\\
p2p-Gnutella31 &	62,586&	147,892& Same as above (at different time)\\
\end{tabular}
\end{ruledtabular}
\end{table}

\begin{table}
\caption{Electronic circuits \cite{milo2002network}}
\begin{ruledtabular}
\begin{tabular}{lrrl}
Name & $N$ & $L$  & description \\\hline
s838 & 512 & 819 & Electronic sequential logic circuit\\
s420 & 252 & 399 & Same as above\\
s208 & 122 & 189 & Same as above\\
\end{tabular}
\end{ruledtabular}
\end{table}

\begin{table}
\caption{World Wide Web (WWW)}
\begin{ruledtabular}
\begin{tabular}{lrrl}
Name & $N$ & $L$  & description \\\hline
stanford.edu \cite{leskovec2014snap}& 281,903 & 1,992,636 & WWW from nd.edu domain \\
nd.edu \cite{albert1999internet} & 325,729 & 1,090,108 & WWW from stanford.edu domain\\
\end{tabular}
\end{ruledtabular}
\end{table}

\begin{table}
\caption{Food webs (FW)}
\begin{ruledtabular}
\begin{tabular}{lrrl}
Name & $N$ & $L$  & description \\\hline
baydry \cite{ulanowicz1998network}&	128&	2,137& FW at Florida Bay, Dry Season \\
 baywet \cite{ulanowicz1998network}&	128	&2,106& FW at Florida Bay, Wet Season \\
 Chesapeake \cite{baird1989seasonal}&	39&	177& FW at Chesapeake Bay Mesohaline Net \\
 ChesLower \cite{hagy2002eutrophication}&	37&	178 & FW at Lower Chesapeake Bay in Summer\\
 ChesMiddle \cite{hagy2002eutrophication}&	37&	209& FW at Middle Chesapeake Bay in Summer\\
 ChesUpper \cite{hagy2002eutrophication}&	37&	215& FW at Upper Chesapeake Bay in Summer \\
 CrystalC \cite{ulanowicz2012growth}&	24&	125& FW at Crystal River Creek (Control) \\
 CrystalD \cite{ulanowicz2012growth}&	24&	100& FW at Crystal River Creek (Delta Temp) \\
 cypdry \cite{Pajek}&	71&	640& FW at Cypress, Dry Season \\
 cypwet \cite{Pajek}&	71&	631& FW at Cypress, Wet Season \\
 Everglades \cite{ulanowicz2000network}&	69&	916& FW at Everglades Graminoid Marshes \\
 Florida \cite{ulanowicz1998network}&	128&	2,106& FW at Florida Bay Trophic Exchange Matrix \\
 gramdry \cite{ulanowicz2000network}&	69&	915& FW at Everglades Graminoids, Dry Season\\
 gramwet \cite{ulanowicz2000network}&	69&	916& FW at Everglades Graminoids, Wet Season \\
 grassland \cite{dunne2002food}&	88&	137& FW at Grassland \\
 littlerock \cite{martinez1991artifacts}&	183&	2,494& FW at Little Rock lake.\\
 mangdry \cite{patricio2005well}&	97&	1,491& FW at Mangrove Estuary, Dry Season \\
 mangwet \cite{patricio2005well}&	97&	1,492& FW at Mangrove Estuary, Wet Season \\
 Maspalomas \cite{almunia1999benthic}&	24&	82& FW at Charca de Maspalomas \\
 Michigan \cite{Pajek}& 39&	221& FW at Lake Michigan Control network \\
 Mondego \cite{patricio2005well}&	46&	400& FW at Mondego Estuary - Zostrea site \\
 Narragan \cite{patricio2005well}&	35&	220& FW at Narragansett Bay Model \\
 Rhode \cite{Pajek}&	20&	53& FW at Rhode River Watershed - Water Budget \\
 seagrass \cite{christian1999organizing}&	49&	226& FW at St. Marks Seagrass.\\
 silwood \cite{memmott2000predators}&	154&	370& FW at Silwood Park\\
 StMarks \cite{baird1998assessment}&	54&	356& FW at St. Marks River (Florida) Flow network\\
 stmartin \cite{goldwasser1993construction}&	45&	224& FW at St. Martin Island\\
 ythan \cite{dunne2002food}&	135&	601& FW at Ythan Estuary\\
\end{tabular}
\end{ruledtabular}
\end{table}

\clearpage

\begin{table}
\caption{Neural network}
\begin{ruledtabular}
\begin{tabular}{lrrl}
Name & $N$ & $L$  & description \\\hline
{\it C.elegans} \cite{watts1998collective} & 297 & 2,148 & Neural network of {\it C.elegans}\\
\end{tabular}
\end{ruledtabular}
\end{table}

\begin{table}
\caption{Transcriptional regulatory networks (TRN)}
\begin{ruledtabular}
\begin{tabular}{lrrl}
Name & $N$ & $L$  & description \\\hline
TRN-Yeast-Babu \cite{balaji2006comprehensive} & 4,441 & 12,864 & Transcriptional regulatory network of {\it S.cerevisiae} \\
TRN-Yeast-Alon \cite{milo2002network} & 688 & 1,078 &Same as above (compiled by different group) \\ 
TRN-EC-RDB64 \cite{gama2008regulondb}& 1,550 & 3,234 & Transcriptional regulatory networkof {\it E.coli}\\
TRN-EC-Alon \cite{milo2002network}& 418 & 519 & Same as above (compiled by different group)\\
\end{tabular}
\end{ruledtabular}
\end{table}

\begin{table}
\caption{Protein-protein interaction (PPI) networks of different organisms \cite{stark2006biogrid}}
\begin{ruledtabular}
\begin{tabular}{lrr}
Name & $N$ & $L$  \\\hline
\textit{Arabidopsis thaliana} (Columbia) & 8,149 & 26,578 \\
\textit{Bos taurus} & 394 & 405  \\
\textit{Caenorhabditis elegans} & 3,941 & 8,641  \\
\textit{Candida albicans} (SC5314)& 370 & 412  \\
\textit{Danio rerio} & 235 & 281  \\
\textit{Drosophila melanogaster} & 8,212 & 47,730  \\
\textit{Emericella nidulans} (FGSC A4) & 64 & 71  \\
\textit{Escherichia coli} (K12/MG1655) & 136 & 127  \\
\textit{Gallus gallus} & 336 & 409  \\
\textit{Hepatitus C Virus} & 112 & 164  \\
\textit{Homo sapiens} & 18,632 & 224,270  \\
\textit{Human Herpesvirus 1} & 139 & 183  \\
\textit{Human Herpesvirus 4} & 223 & 295  \\
\textit{Human Herpesvirus 5} & 71 & 79  \\
\textit{Human Herpesvirus 8} & 140 & 191  \\
\textit{Human Immunodeficiency Virus 1} & 1,028 & 1,786 \\
\textit{Mus musculus} & 8,512 & 22,623\\
\textit{Oryctolagus cuniculus} & 182 & 198  \\
\textit{Plasmodium falciparum} (3D7) & 1,227 & 2,545  \\
\textit{Rattus norvegicus} & 3,313 & 5,357  \\
\textit{Saccharomyces cerevisiae} (S288c) & 6,481 & 333,833  \\
\textit{Schizosaccharomyces pombe} (972h) & 3,963 & 67,796  \\
\textit{Solanum lycopersicum} & 25 & 86  \\
\textit{Sus scrofa} & 87 & 75  \\
\textit{Xenopus laevis }& 479 & 683  \\
\end{tabular}
\end{ruledtabular}
\end{table}

\clearpage

\begin{table}
\caption{Communication networks}
\begin{ruledtabular}
\begin{tabular}{lrrl}
Name & $N$ & $L$  & description \\\hline
Cellphone \cite{song2010limits}&	36,595&	91,826& Call network of cell phone users\\
Email-Enron  \cite{leskovec2014snap}	&36,692&	367,662& Email communication network from Enron\\
Email-EuAll   \cite{leskovec2014snap}&	265,214&	420,045& Email network from a large European research institute\\
Email-epoch \cite{eckmann2004entropy}	&3,188&	39,256&Email network in a university\\
UCIonline \cite{opsahl2009clustering}	&1,899&	20,296& Online message network of students at UC, Irvine\\
WikiTalk \cite{leskovec2014snap}	&2,394,385	&5,021,410&Wikipedia talk network\\
\end{tabular}
\end{ruledtabular}
\end{table}

\begin{table}
\caption{Social networks}
\begin{ruledtabular}
\begin{tabular}{lrrl}
Name & $N$ & $L$  & description \\\hline
Epinions \cite{richardson2003trust}&	75,888&	508,837&Who-trusts-whom network of Epinions.com\\
college student \cite{van2003evolution, milo2004superfamilies}&	32&	96&Social networks of positive sentiment (college students)\\
prison inmate \cite{van2003evolution, milo2004superfamilies}&	67&	182& Same as above (prison inmates)\\
Slashdot-1 \cite{leskovec2014snap}&	77,357&	516,575&Slashdot social network \\
Slashdot-2 \cite{leskovec2014snap}&	81,871&	545,671&Same as above (at different time)\\
Slashdot-3 \cite{leskovec2014snap}&	82,144&	549,202&Same as above (at different time)\\
Slashdot-4 \cite{leskovec2014snap}&	77,360&	905,468&Same as above (at different time)\\
Slashdot-5 \cite{leskovec2014snap}&	82,168&	948,464&Same as above (at different time)\\
Twitter \cite{leskovec2014snap}&	81,306&	2,420,766&Social circles from Twitter\\
WikiVote \cite{leskovec2014snap}&	7,115&	103,689&Wikipedia who-votes-on-whom network\\
Youtube \cite{leskovec2014snap}&	1,134,890&	2,987,624& Youtube online social network\\
\end{tabular}
\end{ruledtabular}
\end{table}

\clearpage

\begin{table}
\caption{Intra-organizational networks}
\begin{ruledtabular}
\begin{tabular}{lrrl}
Name & $N$ & $L$  & description \\\hline
Freemans-1 \cite{freeman1979networkers}&	34&	695&Social network of network researchers\\
Freemans-2 \cite{freeman1979networkers}&	34&	830&Same as above (at different time)\\
Freemans-3 \cite{freeman1979networkers}&	32&	460&Same as above (at different time)\\
Consulting-1 \cite{cross2004hidden}&	46&	879& Social network from a consulting company\\
Consulting-2 \cite{cross2004hidden}&	46&	858& Same as above (different evaluation)\\
Manufacturing-1 \cite{cross2004hidden}&	77&	2,326&Social network from a manufacturing company\\
Manufacturing-2 \cite{cross2004hidden}&	77&	2,228& Same as above (different evaluation)\\
\end{tabular}
\end{ruledtabular}
\end{table}

\clearpage

%
%
%
%
%
%
%
%
%
%
%
%
%
%
%
%
%
%
%
%
%
%
%
%
%
%
%
%
%
%
%
%
%
%
%
%
%
%
%
%
%
%
%
%
%
%
%
%
%
%
%
%
%
%
%
%
%
%
%
%
%
%
%
%
%
%
%
%
%
%
%
%
%
%
%
%
%
%
%
%
%
%
%
%
%
%
%
%
%
%
%
%
%
%
%
%
%
%
%
%
%
%
%
%
%
%
%
%
%
%
%
%
%
%
%
%
%
%
%
%
%
%
%
%
%
%
%
%
%
%
%
%
%
%
%
%
%
%
%
%
%
%
%

%
%
%
%
%
%

%
%
%
%
%
%
%
%
%
%
%

%
%
%
%

%
%
%
%
%
%
%
%

%
%
%
%
%
%
%
%
%
%
%
%
%
%
%
%
%
%
%
%
%
%

%
%
%
%
%
%
%
%
%
%
%

%
%
%
%

%
%
%
%

%
%

%

%
%
%
%
%
%
%
%
%
%
%
%
%
%
%
%
%
%
%
%
%
%
%
%
%
%
%
%
%
%
%
%
%
%
%
%
%
%
%
%
%
%
%
%
%
%
%
%
%
%
%
%
%
%
%
%
%
%
%
%
%
%
%
%
%
%
%
%
%
%
%
%
%
%
%
%
%
%
%

\clearpage

{\bf Acknowledgement.} We thank Shlomo Havlin, Mehran Kardar, Wei Chen, Endre Cs\'{o}ka,
Abhijeet Sonawane, Yandong Xiao, Chuliang Song for
valuable discussions. 
 This work was partially supported by the John
 Templeton Foundation (Award No. 51977) and National Natural Science Foundation of China (Grants Nos. 11505095 and 11374159).

{\bf Author Contributions.} 
Y.-Y.L. conceived and designed the project. 
L.T. performed all the analytical calculations and
extensive numerical simulations, as well as empirical data
analysis. 
All authors analyzed the results. 
L.T. and Y.-Y.L. wrote the manuscript. 
A.B. and D.-N.S. edited the manuscript. 

{\bf Competing Interests}. The authors declare that they have no
competing financial interests.

{\bf Correspondence}. Correspondence and requests for materials
should be addressed to Y.-Y.L.~(yyl@channing.harvard.edu).

\end{document}